\documentclass[acmsmall,screen]{acmart}
\usepackage{multirow}
\usepackage{xspace}
\usepackage{color, colortbl}
\usepackage{xcolor}
\usepackage{enumitem}
\usepackage{balance}
\usepackage[most]{tcolorbox}
\usepackage{wrapfig}
\usepackage{lipsum}
\usepackage{graphicx} % Required for \resizebox
\usepackage{booktabs} % Required for \toprule, \midrule, \bottomrule
\usepackage{pifont}   % Required for \ding{51} and \ding{55}
\usepackage{makecell} % Required for line breaks in table cells
\usepackage[skip=1pt,labelfont=bf]{caption}

\colorlet{punct}{red!60!black}
\definecolor{background}{HTML}{EEEEEE}
\definecolor{delim}{RGB}{20,105,176}
\definecolor{yellow}{RGB}{254, 254, 0}
\definecolor{lightblue}{RGB}{0, 254, 254}
\definecolor{mygray}{gray}{0.98}
\newcommand{\intuition}[1]{
\begin{tcolorbox}[colback=white,boxrule=1pt,top=0pt,bottom=0pt,left=1pt,right=2pt,top=2pt,bottom=2pt]
\em #1
\end{tcolorbox}
}
\newcommand{\mytitle}{PlayCoder\xspace}
\newcommand{\datasetname}{PlayEval\xspace}
\newcommand{\revise}[1]{#1}
\begin{document}

\title{PlayCoder: Making LLM-Generated GUI Code Playable}

\author{Zhiyuan Peng}
\authornote{Zhiyuan Peng and Wei Tao contributed equally to this research.}
\email{pzy2000@sjtu.edu.cn}
\affiliation{%
  \institution{Shanghai Jiao Tong University}
  \city{Shanghai}
  \country{China}
}
\author{Wei Tao}
\authornotemark[1]
\email{wtao@ieee.org}
\affiliation{%
  \institution{LIGHTSPEED}
  \city{Shenzhen}
  \country{China}
}
\author{Xin Yin}
\email{xyin@zju.edu.cn}
\affiliation{%
  \institution{Zhejiang University}
  \city{Hangzhou}
  \country{China}
}
\author{Chenhao Ying}
\email{yingchenhao@sjtu.edu.cn}
\affiliation{%
  \institution{Shanghai Jiao Tong University}
  \city{Shanghai}
  \country{China}
}
\author{Yuan Luo}
\authornote{Yuan Luo and Yiwen Guo are the corresponding authors.}
\email{yuanluo@sjtu.edu.cn}
\affiliation{%
  \institution{Shanghai Jiao Tong University}
  \city{Shanghai}
  \country{China}
}
\author{Yiwen Guo}
\authornotemark[2]
\email{guoyiwen89@gmail.com}
\affiliation{%
  \institution{Independent Researcher}
  \country{China}
}
\renewcommand{\shortauthors}{Peng, Tao, et al.}

\begin{abstract}
Large language models (LLMs) have transformed code generation, 
but their ability to generate code for applications with graphical user interfaces (GUIs), 
particularly games, remains underexplored.
Prior code-generation benchmarks assess correctness using test cases, 
but this is insufficient for GUI applications.
These applications are interactive and event-driven, 
and their correctness depends on stateful behavior over sequences of user actions. 
Consequently, evaluation should account for interaction flows and UI state transitions rather than relying solely on pass or fail test outcomes.
To explore the performance of LLMs on GUI applications, 
we construct PlayEval, 
a repository-aware evaluation dataset from \revise{43 multilingual 
(Python, TypeScript, and JavaScript) GUI applications. 
Different from existing GUI benchmarks which are difficult to transplant to Desktop platform, 
PlayEval consists of 6 major categories of GUI applications 
and directly supports evaluation on code generation tasks.}
To enable more reliable assessment beyond simple execution and unit tests, we propose Play@k, 
which measures whether at least one of k generated candidates yields an application that can be played end-to-end without logical errors.
We further develop an LLM-based agent, PlayTester, 
that automates interactive evaluation by driving the GUI through task-oriented playthroughs and checking for logic violations. 
Through systematic evaluation, we demonstrate that 10 state-of-the-art code LLMs struggle to generate logically correct GUI applications, 
achieving near-zero Play@3 scores despite high compilation rates.
To address these, we introduce \mytitle{}, a multi-agent, 
repository-aware framework that writes, evaluates and refines GUI application code via closed-loop control. 
\mytitle{} substantially improves functional correctness and semantic alignment for both open-source and closed-source models, 
achieving up to 38.1\% Exec@3 and 20.3\% Play@3. 
Case studies show that it detects silent logic flaws missed by traditional metrics and repairs them through targeted edits. 
These results indicate that coupling an end-to-end GUI testing agent with repository-aware automated program repair is an effective path towards reliable GUI code generation.
Our implementation is publicly available at \url{https://github.com/Tencent/PlayCoder}.
\end{abstract}




    \maketitle

\section{Introduction}
Large language models (LLMs) have revolutionized software engineering tasks (e.g., code generation and bug fixing), 
achieving impressive results on established benchmarks like HumanEval~\cite{chen2021evaluating}, CoderEval~\cite{yu2024codereval}, 
and SWE-Bench~\cite{jimenez2024swe}. 
These established benchmarks primarily target well-specified, 
self-contained programming tasks amenable to unit test verification. 
Consequently, they inadequately represent the complexities of open-ended environments that require capabilities 
beyond single-shot function synthesis (e.g., multi-step interaction with live systems, 
sustained stateful execution, and external tool integration). 
Such capabilities are fundamental to interactive graphical user interface (GUI) applications. 
GUI code generation presents distinct evaluation challenges: 
models must handle event-driven control flow, persistent and evolving application state, 
and complex user interaction patterns. While algorithmic tasks permit assessment through input-output validation, 
GUI systems necessitate interactive verification procedures that existing evaluation paradigms cannot accommodate.
This evaluation limitation becomes particularly acute because GUI applications frequently manifest silent behavioral failures, 
where syntactically correct and executable code violates fundamental application logic.

Current evaluation frameworks inadequately assess the behavioral requirements of GUI applications. 
While SWE-Bench~\cite{jimenez2024swe} advances repository-aware evaluation, 
it relies primarily on unit tests and static analysis. 
These approaches prove insufficient for GUI applications, 
where behavioral correctness cannot be captured through traditional test cases~\cite{bian2025you}.
\revise{While web-based GUI testing frameworks (e.g., Selenium, Playwright) automate interaction through DOM manipulation, 
they fundamentally rely on accessible structural representations that many GUI applications lack. 
Canvas-based applications, desktop games, and native GUI programs render content 
directly to pixels without exposing DOM trees or accessibility APIs, 
making structure-based testing infeasible. }
GUI applications may pass unit tests yet exhibit interactive failures 
that manifest only during runtime execution.
As shown in Fig.~\ref{fig:motivation}, 
consider a Flappy Bird game that compiles without errors 
but allows the bird to pass through obstacles, 
violating core game mechanics while producing no exceptions.
Such failures remain undetectable through unit tests because obstacles are randomly generated with non-deterministic coordinates, 
making comprehensive test case coverage impractical.
Consequently, developers typically rely on human testers to identify and report behavioral bugs, 
a process that is both time-consuming and costly.
This gap demonstrates that current approaches fundamentally struggle with interactive GUI applications.

To address these, we establish evaluation benchmark, metrics, and method for GUI-based code generation.
We develop a GUI-testing methodology that capture behavioral correctness through automated user interaction simulation.
These methodologies are complemented by \datasetname{}, a curated benchmark of \revise{43 diverse multilingual (Python, TypeScript, and JavaScript) GUI applications spanning six major categories 
(e.g., classic games, MMORPG games, productivity tools)} with verifiable GUI behaviors. \revise{We focus on interactive GUI applications. Games are selected as examples to represent the challenge of interactivity: requiring frequent state updates, event handling, and operating solely based on visual feedback (unlike web-GUIs with accessible DOM structures). The benchmark includes general-purpose desktop-applications because it can reflect common use cases and evaluate the generality of such less complex, yet challenging applications.}
We propose Play@k, a behavioral correctness metric that measures whether generated code can be interactively executed end-to-end without logical errors.
Since Play@k requires code first to pass unit tests before GUI validation, it provides a more stringent assessment than traditional Pass@k metrics.

Building upon these evaluation foundations, we propose \mytitle{}, 
a multi-agent framework that leverages our evaluation methods for robust GUI-based code generation.
\revise{The framework employs two specialized agents: a repository-aware coding agent (PlayDeveloper) for initial code generation
and an automated program repair agent (PlayRefiner) that iteratively refines code based on evaluation feedback from PlayTester.
PlayTester serves as the behavioral testing framework that verifies correctness across programming languages and platforms (Windows, macOS, and X11-based Linux distributions).}
As shown in Figure~\ref{fig:approach}, 
PlayDeveloper generates initial code, then PlayTester detects behavioral deviations,
enabling PlayRefiner to autonomously debug and modify code through successive test-repair cycles.
This iterative loop helps produce code that is syntactically valid and better aligned with the requirements.
\revise{Iterative refinement~\cite{kim2023language,madaan2023self,yang2024swe,tang2024code} is prevalent. However, PlayCoder differs in \textit{what-drives-the-loop}:
1) visual vs. textual feedback: Prior works rely on \textbf{textual signals}. PlayCoder closes the loop using \textbf{visual feedback} (i.e., screenshots) and \textbf{dynamic interaction} (e.g., mouse or keyboard operation), enabling it to fix ``silent failures'' (e.g., invisible text, unresponsive buttons) that text-based oracles miss.
2) active exploration vs. passive testing: Standard loops use pre-defined test suite. Employing PlayTester dynamically explores the UI to discover bugs.}
Our evaluation demonstrates \mytitle{}'s effectiveness across \revise{multilingual} GUI applications.
Using GPT-5-mini, our framework achieves 26.8\% Exec@3 and 9.8\% Play@3, 
compared to the best baseline (DeepCode) with 17.9\% Exec@3 and 6.4\% Play@3.
With Claude-Sonnet-4, \mytitle{} reaches 36.8\% Exec@3 and 20.3\% Play@3, 
demonstrating model-agnostic benefits and establishing a new paradigm for interactive GUI application code generation.
Our main contributions are summarized as follows:

\begin{itemize}[leftmargin=*]

    \item We present a comprehensive benchmark for GUI application code generation, consisting of the \datasetname dataset and the Play@k metric. Covering 43 multilingual applications across 6 domains (e.g., MMORPGs), our experiments uncover a severe performance gap in behavioral correctness. With the top-performing model achieving only 9.9\% Play@3 (18.6\% Exec@3) and the weakest baseline scoring <1\%, our findings highlight critical challenges that current methods fail to solve.
    
    \item We propose PlayTester, 
    a GUI Testing Agent that serves dual purposes: 
    As an evaluator, 
    PlayTester detects subtle behavioral failures (e.g., collision detection errors, 
    event handling inconsistencies) that traditional unit tests overlook.
    As a feedback provider, 
    PlayTester provides precise behavioral diagnostics to guide iterative code improvement, 
    addressing repository hallucination where models generate syntactically correct but behaviorally incorrect code.
    \item We propose \mytitle{}, 
    a novel multi-agent framework that integrates repository-aware code generation, 
    automated GUI behavioral testing, and iterative program repair.
    \revise{The framework employs two specialized agents (PlayDeveloper and PlayRefiner)
    that collaborate through structured test\&repair cycles, using visual feedback from the behavioral evaluation framework.} 
    \mytitle{} achieves up to 20.2\% higher Exec@3 and 11.0\% higher Play@3 compared to baselines, 
    with consistent effectiveness across diverse LLM architectures and superior \revise{cost-effectiveness}.
\end{itemize}

\section{Motivation}
\subsection{A Motivating Example}

\begin{wrapfigure}{r}{0.475\textwidth}
    \centering
    \includegraphics[width=0.5\textwidth]{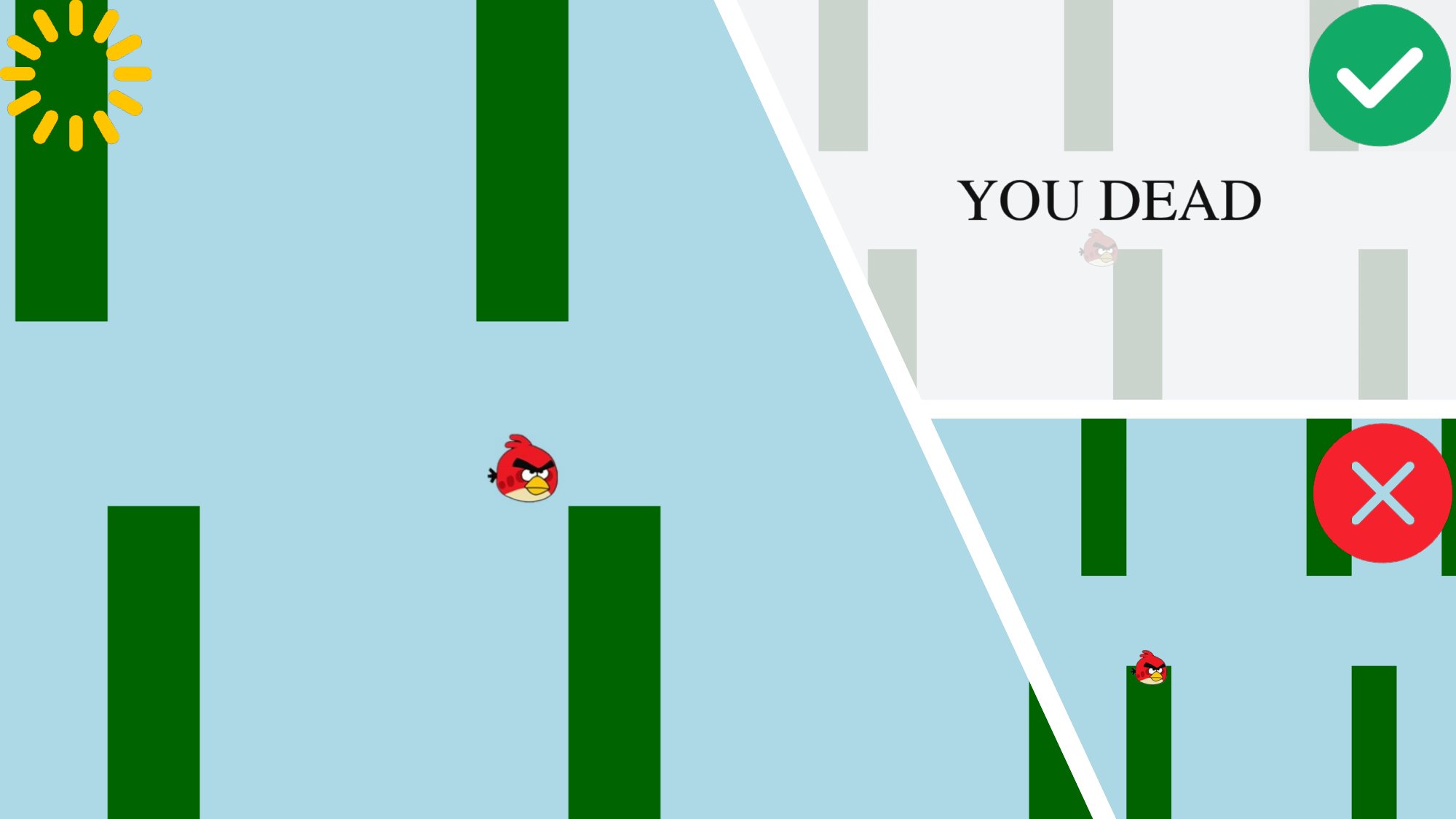}
    \caption{\textit{Flappy Bird} generated by GPT-4o-mini and human programmer. 
    Top-right: code written by a human programmer, where collision correctly kills the bird and ends the game.
    Bottom-right: code generated by GPT-4o-mini, where the bird can pass through the pipe, which is a critical logic flaw.
    }
    \label{fig:motivation}
\end{wrapfigure}

Fig.~\ref{fig:motivation} illustrates an example in \textit{Flappy Bird} generated by GPT-4o-mini and a human programmer. 
The program compiles and runs, but it allows the bird to pass through obstacles, so the game never ends (as shown in the bottom-right of Fig.~\ref{fig:motivation}). 
In the correct behavior, a collision should kill the bird and terminate the game (as shown in the top-right of Fig.~\ref{fig:motivation}).
Such failures do not raise exceptions or cause crashes, allowing them to slip past evaluations that only verify compilation or test cases.

\noindent \textbf{Challenge 1: Testing Dilemma for GUI Application Code Generation.}
Traditional evaluation of code emphasizes compilation success and unit test, which are inadequate for GUI applications. 
Unlike function-level code validated by input–output pairs, GUI programs demand interactive, stateful, and temporal validation that current paradigms miss~\cite{bian2025you}. 
In the \textit{Flappy Bird} example, the code runs without runtime errors, yet critical behavioral flaws (e.g., the bird penetrating pipes) remain undetected. Metrics (e.g., Pass@k) cannot distinguish a GUI application from one with broken logic, calling for behavioral testing via interactive execution.

\noindent \textbf{Challenge 2: Insufficient Benchmarks for GUI Application Code Generation.}
Existing code generation benchmarks systematically underrepresent GUI application generation (including GUI-based games), despite its prevalence in practice. 
HumanEval~\cite{chen2021evaluating} and CoderEval~\cite{yu2024codereval} focus on algorithmic or function-level tasks; while SWE-Bench~\cite{jimenez2024swe} advances repository-aware evaluation, it fundamentally relies on unit test passage and static analysis, making it insufficient for GUI applications where behavioral correctness requires interactive validation that traditional test cases cannot capture. 
Preliminary experiment from \datasetname{} (i.e., Table~\ref{tab:rq1}) shows a sharp degradation from executability to behavioral validity: the best model (Claude-Sonnet-4) achieves 18.6\% execution correctness but only 9.9\% behavior correctness on Python; GPT-5 drops from 17.5\% to 6.9\% \revise{(refer to Section~\ref{sec:preliminary} for more details)} on Python. 
This gap yields a false sense of competence: models 
that score highly on traditional benchmarks perform poorly on high-complexity GUI tasks involving event handling, 
state updates, and physics-based animation.
\noindent \textbf{Challenge 3: \revise{Difficulty} in Repository-Aware GUI Code Generation.}
While recent repository-aware code generation methods (e.g., MetaGPT~\cite{hong2024metagpt}, DeepCode~\cite{deepcode2025}) 
have made progress in incorporating repository context, 
they still face significant challenges when applied to GUI application code generation.
These methods excel at retrieving relevant code snippets and API documentation, 
yet they struggle with GUI-specific behavioral correctness because traditional repository analysis focuses on 
syntactic patterns rather than interactive semantics.
Observable behavioral failures, such as the bird in \textit{Flappy Bird} passing through pipes without collision detection, 
often occur despite syntactically correct API usage and proper imports.
GUI applications require understanding of event loops, state transitions, 
and temporal properties that cannot be captured through static repository analysis alone.
Even with comprehensive repository grounding, existing agents fail to distinguish between code 
that compiles correctly and code that behaves correctly during interactive execution, 
leading to systematic behavioral hallucinations in GUI contexts. \revise{In addition to this, 
existing tools (e.g., Selenium) and recent LLM-based testers (e.g., GPTDroid~\cite{liu2023chatting}, 
LLMDroid~\cite{wang2025llmdroid}, VETL~\cite{wang2024leveraging}) focus on \textbf{Web or Mobile} platforms. 
They rely on structured DOMs or accessibility trees that are missing or non-standard in desktop Python GUIs 
(especially PyGame, which renders to a canvas). Transplanting them to support desktop GUI applications is impractical.}

\subsection{Key Ideas}
Based on the above challenges, we present three complementary ideas to enable and evaluate repository-aware GUI application code generation with behavioral guarantees.

\noindent \textbf{Key Idea 1: Overlooked GUI Generation Challenges.}
Our preliminary experiments on \datasetname{} reveal that GUI application code generation poses 
significantly greater challenges than previously recognized, with systematic failures overlooked 
by existing evaluation paradigms.
Pass@k, which relies solely on unit tests, 
fails to capture GUI-specific errors such as incorrect collision detection, 
broken event handling, or randomly generated map inconsistencies that appear functionally correct but render applications unusable.
Current benchmarks inadequately represent GUI application complexity, 
while existing methods lack proper assessment frameworks for interactive behavioral correctness.
Preliminary studies in Section~\ref{sec:preliminary} demonstrate substantial performance degradation across state-of-the-art models: 
Claude-Sonnet-4 drops from 18.6\% Exec@3 to 9.9\% Play@3, 
while GPT-5 plummets from 17.5\% Exec@3 to merely 6.9\% Play@3, 
revealing a critical gap between syntactic correctness and behavioral validity that traditional metrics systematically miss.

\noindent \textbf{Key Idea 2: Novel GUI Benchmark with Hierarchical Behavioral Testing.}
We contribute \datasetname{}, a comprehensive repository-aware benchmark that advances GUI application code generation 
evaluation through three key innovations: (1) a curated dataset of diverse GUI applications (e.g., GUI-based games, productivity tools) with repository scaffolds and behavioral specifications; 
(2) a novel evaluation methodology that introduces hierarchical testing with our PlayTester; 
and (3) a rigorous metric framework that detects subtle behavioral failures overlooked by traditional approaches.
Our Play@k metric represents a significant methodological advancement: 
it exclusively evaluates code that first passes unit tests (Pass@k), 
then subjects it to automated GUI behavioral testing via our specialized PlayTester, 
making it substantially more stringent than existing metrics.
This approach ensures Play@k captures hidden defects 
(e.g., collision detection failures, event handling inconsistencies, temporal property violations) 
that appear correct under unit testing but manifest as critical behavioral flaws during interactive execution.
\datasetname{} emphasizes event handling, state management, and physics/animation with standardized interaction scripts, 
enabling fair and reproducible model comparison under identical behavioral testing conditions.

\noindent \textbf{Key Idea 3: Multi-Agent Framework with PlayTester for Hallucination Mitigation.}
We design a multi-agent framework in which PlayTester serves \emph{both} as a rigorous behavioral evaluator \emph{and} as a hallucination-reduction feedback source for GUI code generation. 
Beyond its evaluation role, 
the PlayTester provides behavioral feedback that enables collaborative agents to 
systematically address repository hallucination, 
a pervasive issue where models generate syntactically correct but behaviorally incorrect code.
The framework integrates retrieval-augmented generation, GUI testing, 
and automated program repair (APR) in a closed-loop control via a multi-agent system where PlayTester acts as a behavioral oracle.
PlayDeveloper generates repository-aware code while invoking tools to retrieve context from the codebase.
PlayTester executes applications and provides precise behavioral diagnostics 
(e.g., collision detection failures, event handling inconsistencies, state transition violations) that guide subsequent generation attempts.
PlayRefiner performs targeted repairs using these behavioral insights: 
(1) compilation errors trigger \texttt{Validator} invocations; 
(2) API misuse and undefined symbols prompt context-guided adaptations via \texttt{ContextSearchTool}; 
(3) behavioral failures (e.g., incorrect physics) trigger logic adjustments based on PlayTester feedback.
This approach transforms the PlayTester from a passive evaluator into an active participant that reduces hallucination by providing actionable behavioral signals throughout the generation process.

\revise{\section{Benchmark and Evaluation}}
\label{sec:experiment}
\noindent
\revise{In this section, we introduce the benchmark construction, baselines, 
evaluation method, and preliminary studies.}
\revise{For all experiments, we generate \(n = 3\) samples per problem and 
compute the unbiased estimator with \(k \leq 3\). 
All baselines are evaluated with their official configuration. 
We mitigate the influence of environmental factors 
(e.g., hardware differences) by providing standardized configurations and 
conducting all experiments with 5 repetitions.}

\subsection{Benchmark Construction}
\label{sec:dataset}
Following established best practices for code generation benchmarks~\cite{khan2023xcodeeval,yan2023codescope,yu2024codereval,li2024deveval,peng2025soleval}, 
we present \datasetname, a repository-aware benchmark for GUI application code generation in Python, TypeScript, and JavaScript.  

\subsubsection{Selection Criteria}
\label{sec:select_criteria}
We curated repositories using the following criteria: 
(1) \textit{Historically Active Development}: \revise{repositories with commits within the past 12 months at the time of initial selection, or demonstrated sustained development history ($\geq$6 months of active maintenance) and achieved feature completeness before archival;} 
(2) \textit{Community validation}: most projects with >100 GitHub stars 
(We also include projects with $\leq$ 100 stars but have excellent deployability and representative of a certain category); 
(3) \textit{Functional completeness}: applications that demonstrate complete GUI workflows 
rather than isolated snippets; (4) \textit{Framework diversity}: 
coverage of major Python GUI frameworks, including PyQt, PySide, Tkinter, and Pygame; 
(5) \textit{Exemplary value}: projects with clear structure and documentation suitable for code 
generation evaluation. \revise{We selected non-trivial functions central to the app's logic 
(e.g., game-loop, event-handlers) rather than utility helpers. We apply a filter to focus extraction on behavior‑rich code: after excluding docstrings and decorator lines, we use a default threshold min\_lines = 28 because we empirically find that functions with fewer than 28 lines are more likely to be utility helpers, simple accessors and rarely implement core interactive behaviors (e.g., game‑loops or event‑handlers).}

\begin{table}[htbp!]
  \centering
  \caption{Comprehensive code complexity statistics of \datasetname across categories. 
  CC = Cyclomatic Complexity (avg per file), ND = Nesting Depth (avg levels), 
  CF/kLOC = Control Flow structures per 1000 LOC. 
  \revise{LOC reports total lines of code across all projects.}}
  \label{tab:dataset-statistics}
  \resizebox{\linewidth}{!}
  {
  \begin{tabular}{lcccccccccc}
  \toprule
  \textbf{Category} & \textbf{Projects} & \textbf{Files} & \textbf{LOC} & \textbf{Functions} & \textbf{Classes} & \textbf{CC$\uparrow$} & \textbf{ND$\uparrow$} & \textbf{CF/kLOC$\uparrow$} & \textbf{LOC/Func$\uparrow$} & \textbf{Test Cases$\uparrow$} \\
  \midrule
  Game Emulation & 1 & 89 & 26,699 & 1,350 & 289 & 8.2 & 11.4 & 24.1 & 19.7 & 86 \\
  Classic Games & 6 & 34 & 2,605 & 72 & 19 & 6.7 & 8.8 & 19.4 & 36.2 & 24 \\
  Game Engine & 1 & 43 & 12,484 & 661 & 87 & 9.5 & 13.2 & 28.3 & 18.8 & 27 \\
  Standalone Applications & \revise{24} & \revise{387} & \revise{123,442} & \revise{2,069} & \revise{172} & \revise{10.8} & \revise{10.8} & \revise{32.1} & \revise{55.7} & 1,539 \\
  Desktop Widgets & 9 & 67 & 21,420 & 314 & 24 & 11.8 & 12.8 & 31.4 & 68.2 & 396 \\
  \revise{MMORPG Games} & \revise{2} & \revise{17} & \revise{1,782} & \revise{31} & \revise{4} & \revise{8.8} & \revise{3.4} & \revise{21.9} & \revise{57.5} & \revise{32} \\
  \midrule
  \revise{\textbf{Total}} & \revise{\textbf{43}} & \revise{\textbf{637}} & \revise{\textbf{188,432}} & \revise{\textbf{4,497}} & \revise{\textbf{595}} & \revise{\textbf{10.2}} & \revise{\textbf{11.0}} & \revise{\textbf{30.4}} & \revise{\textbf{40.0}} & \revise{\textbf{2,104}} \\
  \bottomrule
  \end{tabular}
  }
\end{table}

\subsubsection{Dataset Composition} 
As shown in Table~\ref{tab:dataset-statistics}, 
\datasetname comprises \revise{43 diverse GUI applications} including GUI-based games, productivity tools, multimedia applications, etc.
\revise{The benchmark covers three programming languages: Python, TypeScript, and JavaScript. We selected Python as it is popular in AI/ML research and has rich GUI-bindings. Furthermore, most baselines (except OpenManus) are optimized only for Python, making it a ``common supported language''. We included TypeScript and JavaScript because, according to GitHub's 2025 Octoverse report\footnote{\url{https://github.blog/news-insights/octoverse/octoverse-a-new-developer-joins-github-every-second-as-ai-leads-typescript-to-1}}, TypeScript became the most widely used programming language on GitHub and now serves as the default scaffold language for most mainstream frontend frameworks.}
\datasetname encompasses six major categories: 
(1) \textit{Game Emulation} comprising a complete Game Boy emulator (i.e., PyBoy) with sophisticated hardware simulation capabilities; 
(2) \revise{\textit{Classic Games} including traditional arcade-style games (i.e., 2048, Snake, Flappy Bird, Sudoku, Chrome Dragon) and strategy-based games like Solitaire and Chess variants;} 
\revise{(3) \textit{MMORPG Games} featuring two high-star TypeScript games (both with >1000 stars), CyberCodeOnline and biomes-game, to evaluate cross-language capability;}
(4) \textit{Game Engine} featuring the Jupylet framework for educational game development; 
(5) \revise{\textit{Standalone Applications} encompassing general-purpose applications implemented in Python, JavaScript, and TypeScript that can be further categorized into productivity tools (e.g., text editors, file managers), multimedia applications (e.g., media players), and web-based applications (e.g., Spotify client, Windows 11 simulator). This category also includes small-scale applications like calculators.}
(6) \textit{Desktop Widgets} \revise{comprising interactive components (e.g., color pickers, range sliders). The classification distinction relies on window resizability: Standalone Applications typically support dynamic window resizing, whereas Desktop Widgets are often fixed-size components of a larger user interface;}

\subsubsection{Benchmark Structure} 
\begin{figure}
    \vspace{-0.2cm}
    \centering
    \includegraphics[width=0.88\linewidth]{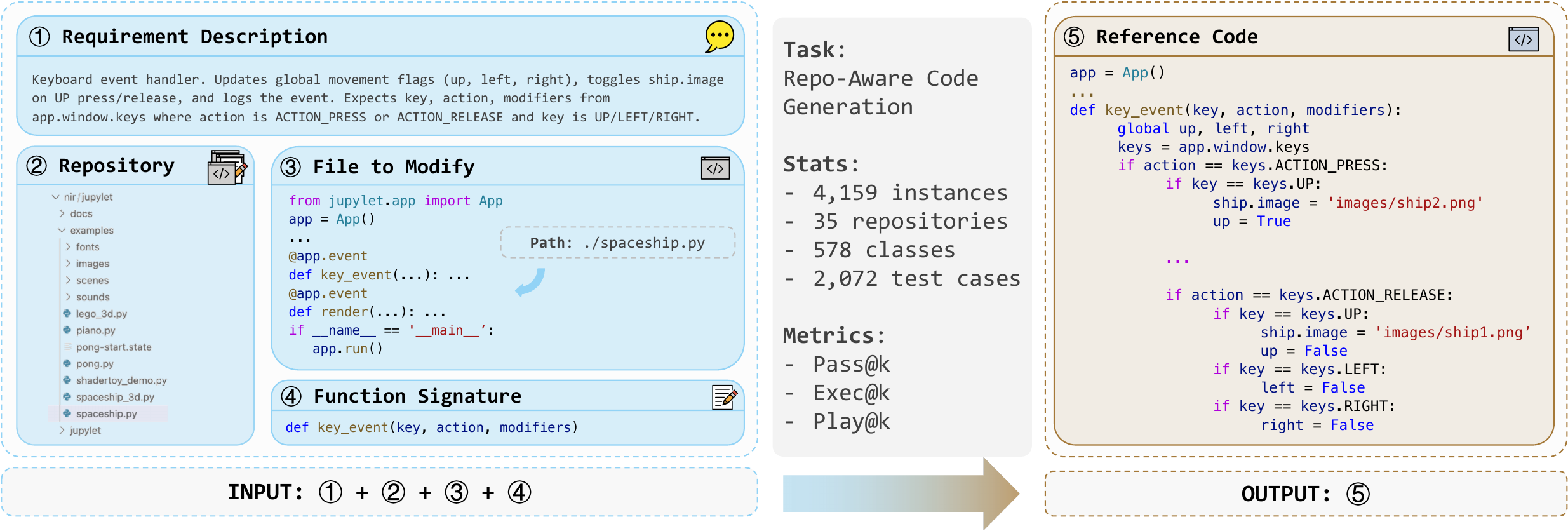}
    \caption{The structure of \datasetname Data.}
    \label{fig:dataset}
    \vspace{-0.2cm}
\end{figure}

As shown in Fig.~\ref{fig:dataset}, \datasetname uses three evaluation metrics. 
The benchmark aims at repository-aware code generation, where each evaluation instance comprises: (1) \textbf{Function Signature} extracted from the original codebase, providing the exact method declaration with parameter types and return specifications; (2) \textbf{Requirement} automatically generated from the original function body using LLM-based docstring generation, which analyzes implementation logic to produce concise natural language descriptions of the function's purpose, behavior, parameters, and expected outcomes; (3) \textbf{Repository Context} containing relevant import statements, class definitions, and related function bodies from the same codebase to enable repository-aware code generation. \revise{The context for PlayCoder is the same as for human developers. To achieve it, we revert the repository to a certain state by \texttt{`git checkout'}.}
\revise{Requirements were generated by GPT-4o-mini and manually verified on a subset ($\approx 10\%$) by 3 developer experts. Through a voting strategy, over 95.6\% of the requirements were deemed high-quality. The original repository's unit tests serve as ground truth (line coverage: 47.2\%, branch coverage: 32.1\%), reflecting the inadequate test coverage of real-world projects. Therefore, Play@k is critical for interactive validation.}

\subsubsection{Evaluation Workflow} 
\label{sec:evaluation_workflow}
The evaluation pipeline proceeds through three stages:

\textit{Compilation and Execution Stage:} Generated functions undergo Python compilation testing to measure \textit{Exec@k}, the percentage of problems for which at least one solution among k samples executes successfully without runtime errors, syntax errors, or import failures. This metric evaluates basic code correctness and syntactic validity.

\textit{Unit Testing Stage:} Functions that pass compilation are evaluated against comprehensive test suites to measure \textit{Pass@k}, 
the percentage of problems for which at least one solution among k samples passes all provided unit tests. 
Due to the lack of sufficient test cases in the original projects, we automatically generate a more robust suite. 
These test cases are created using an LLM-based analysis of the original function implementations, 
covering unit tests, integration tests, functional tests, and edge cases with proper mocking and isolation strategies.

\textit{Behavioral GUI Testing Stage:} For GUI applications, 
our specialized GUI Behavioral Testing performs interactive validation to measure \textit{Play@k}, 
the percentage of problems for which at least one solution among k samples demonstrates correct behavioral semantics in live application environments. 
\revise{This testing is conducted using PlayTester (detailed in Section~\ref{sec:playtester}), which performs automated GUI interaction and validation.}
For games with explicit objectives (e.g., winning conditions), 
the testing strategy focuses on achieving game completion through strategic gameplay. 
For general GUI applications, 
the testing approach emphasizes comprehensive feature coverage through carefully curated interaction sequences. 
\revise{\textit{Note that Exec@k and Pass@k are deterministic metrics, 
which are not affected by the reliability of LLM backbones.}} 

\subsubsection{Advanced Complexity Analysis} 
The benchmark exhibits a fine-grained complexity profile that stresses modern code generation models along multiple dimensions:

\textit{Cyclomatic Complexity:} The dataset shows an average cyclomatic complexity of 10.2 per file, 
with GUI applications exhibiting the highest complexity (12.6) due to extensive event handling and user interaction logic. Game-related projects maintain moderate complexity (i.e., 6.5 to 9.5), reflecting focused algorithmic implementations, while the overall distribution ranges from simple utilities to sophisticated emulation systems.

\textit{Nesting and Control Flow:} The benchmark exhibits an average nesting depth of 11.0 levels, with 107 files whose \emph{per-file maximum} nesting depth exceeds 20 levels, creating substantial structural complexity. Control-flow analysis reveals 4,480 conditional statements (e.g., if statements), 814 loops (e.g., for statements), and 220 exception-handling blocks, yielding a control-flow density of 30.4 structures per 1000 lines of code, which is substantially higher than typical code generation benchmarks.

\subsection{Baselines}
\label{sec:baselines}
We evaluate \mytitle against: (i) state-of-the-art LLMs and (ii) advanced LLM-based approaches.
\revise{Our baseline selection follows two key principles: (1) breadth of model capabilities, spanning general-purpose and coding-specialized LLMs from both open-source and closed-source families, and (2) coverage of representative LLM-based enhancement strategies, including widely-recognized prompt-based and agentic approaches proven effective for code generation tasks.}

\subsubsection{Basic LLMs}
We consider ten state-of-the-art (SOTA) LLMs spanning diverse architectures, model families, and parameter scales.
\revise{This selection encompasses both general-purpose LLMs (GPT-5, GPT-5-mini, GPT-4o, GPT-4o-mini, Claude-Sonnet-4, Claude-Sonnet-3.7, Grok-3-mini, GLM-4.5, DeepSeek-V3) and coding-specialized models (Qwen3-Coder), ensuring comprehensive coverage of different optimization objectives.}
The suite ranges from compact, efficient models to large-scale systems and includes both closed- and open-source variants with varying degrees of code specialization.
For closed-source models, we choose GPT-5, GPT-5-mini, GPT-4o, GPT-4o-mini, Claude-Sonnet-4, Claude-Sonnet-3.7, and Grok-3-mini.
For open-source models, we include Qwen3-Coder (480B), GLM-4.5 (355B), and DeepSeek-V3 (671B).

\subsubsection{LLM-based Approaches}
Beyond basic LLMs, we evaluate state-of-the-art LLM-based approaches that incorporate repository-aware retrieval, structured reasoning, or specialized code-generation capabilities.
\revise{Our selection prioritizes methods with demonstrated effectiveness in code generation research and practical applicability to GUI application development:}
\begin{itemize}[leftmargin=*]
    \item \textbf{SCoT}~\cite{li2025structured}: a widely-cited prompt-based approach using structured chain-of-thought prompting with an 8-step reasoning pipeline for code generation;
    \item \textbf{HCPCoder}~\cite{zhang2025hierarchical}: Hierarchical context pruning with repository-aware prompt generation, retrieving code examples and import patterns via semantic similarity;
    \item \textbf{MetaGPT}~\cite{hong2024metagpt}: \revise{a popular agentic framework (2051 citations) for code generation that simulates} software development processes through specialized roles\revise{; the original paper~\cite{hong2024metagpt} demonstrates its suitability for generating games (e.g., 2048, Snake), making it particularly relevant to our benchmark};
    \item \textbf{OpenManus}~\cite{openmanus2025}: General-purpose multi-agent framework for complex task execution;
    \item \textbf{DeepCode}~\cite{deepcode2025}: AI-powered development platform that automates code generation and implementation; the multi-agent system translates requirements into functional code.
\end{itemize}

\subsection{PlayTester}
\label{sec:playtester}

PlayTester implements multi-modal testing capabilities through three specialized components that collaborate to validate GUI application behavior, \revise{of which the prompts are specified in~\cite{replication}.}

\subsubsection{Visual Observer Module}
The \texttt{VisualObserver} captures application state via screenshots using \texttt{pyautogui} and \texttt{PIL}. 
It supports region-specific capture and performs window detection using platform-specific APIs (e.g., AppleScript on macOS, Win32 on Windows).
\revise{The module caches recent frames (specifically the last three screenshots to differentiate animations from static states) 
and provides image comparison for state change detection. 
Screenshots are captured after each action execution 
(i.e., one second after the action is completed).
The \texttt{VisualObserver} supports Windows, macOS, and X11-based Linux distributions. 
Wayland-based systems are excluded because Wayland's security architecture prevents cross-window screenshot capture and input injection via standard APIs. 
And support for Wayland-based systems is still under PR for pyautogui\footnote{\url{https://github.com/asweigart/pyautogui/pull/936}}.
}

\subsubsection{Action Executor Module}
The \texttt{ActionExecutor} translates test strategies into specific GUI operations: 
\texttt{click(x, y)}, \texttt{type(text)}, \texttt{hotkey(keys)}, 
\texttt{press(key)}, \texttt{scroll(x, y, direction)}, \texttt{wait(duration)}, 
and \texttt{finish(success/failure)}. 
The module includes safety mechanisms (e.g., coordinate boundary checks 
and failsafe cursors) and maintains execution history for debugging. 
Actions are parsed from structured LLM output using the \texttt{ActionParser}.
\subsubsection{Test Manager}
The \textit{Test Manager} integrates vision-language models to plan tests. 
It processes screenshots and textual context to generate strategies using specialized prompts for GUI analysis, 
test strategy generation, and action decision-making.
\revise{The agents in all phases except for test strategy generation have one pre-defined prompt-template. 
We use two test-strategy prompt templates because GUI applications exhibit two distinct interaction regimes.
Games typically have explicit objectives and terminal conditions (e.g., win/lose states, scores),
so effective testing is goal-driven and focuses on reaching completion-critical states.
In contrast, non-game applications often lack a natural terminal state, where effective testing
is coverage-driven and emphasizes traversing UI workflows (e.g., menus) to maximize feature coverage. The behavioral testing phase is fully automated during verification. Tests are reusable across consistent screen resolutions.}

\subsubsection{Case Study}
We illustrate PlayTester's evaluation approach using a representative 2048 implementation, 
focusing on how it perceives visual state, plans interactions, and verifies behavioral properties while maintaining coherent gameplay progression.
As shown in Fig.~\ref{fig:example}, 
the tester simultaneously validates game functionality and maintains strategic progression in an early-game state (i.e., play and test simultaneously). 
Three key capabilities are highlighted:

\begin{figure}[t]
    \vspace{-0.2cm}
    \centering
    \includegraphics[width=0.88\textwidth]{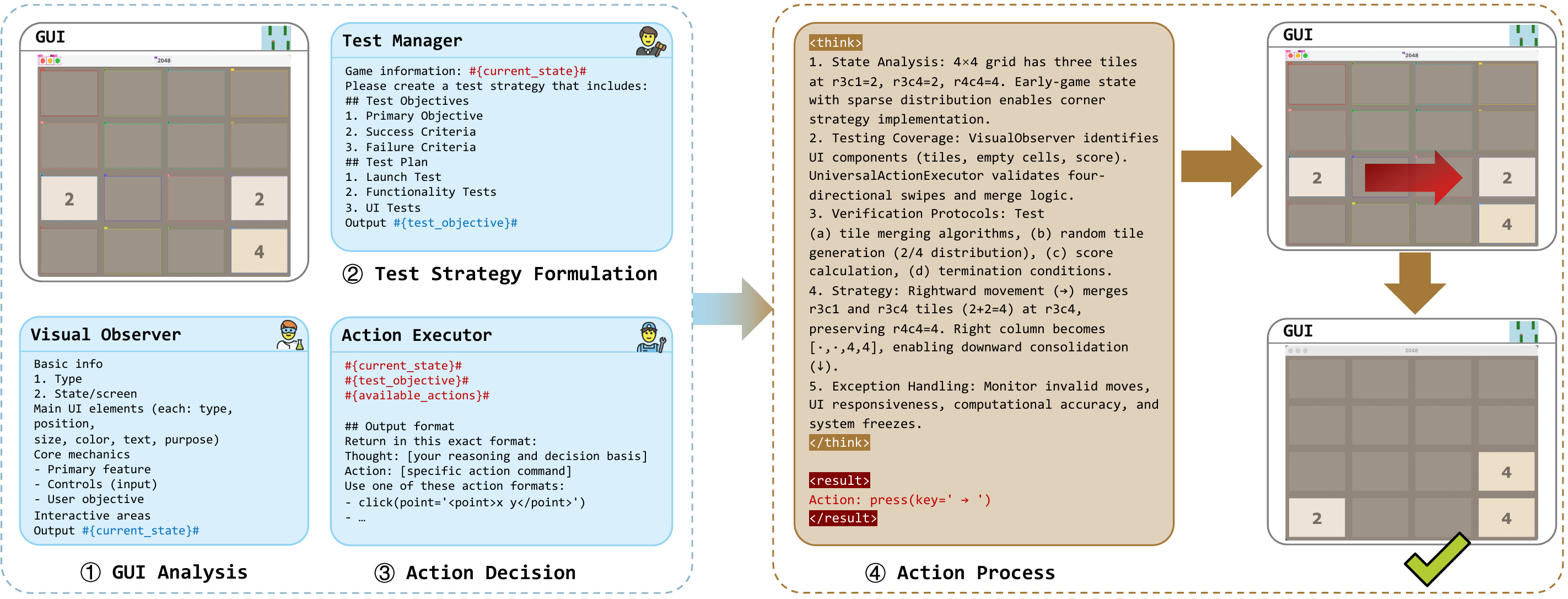}
    \caption{Testing a 2048 implementation. Left: the rendered $4\times4$ grid shows tiles at (3,1), (3,4), and (4,4) with values 2, 2, and 4. Right: the agent's structured reasoning process including state analysis, coverage assessment, verification protocols, strategy selection, and exception-aware considerations that lead to a recommended rightward swipe.}
    \label{fig:example}
\end{figure}

\begin{itemize}[leftmargin=*]
    \item \textbf{Perception and State Extraction.} The \texttt{VisualObserver} captures the current screen and extracts structured state (e.g., r3c1=2, r3c4=2, r4c4=4; score=8), recognizing a sparse early-game configuration.
    \item \textbf{Adaptive Strategy with Dual Objectives.} Based on real-time analysis, the agent selects a rightward swipe ($\rightarrow$) that advances strategy (merging two 2-tiles into a 4 at r3c4) and validates key mechanics: swipe responsiveness, merge algorithm correctness, and score updates. This achieves coverage while preserving plausible gameplay.
    \item \textbf{Exception-Aware Validation.} The agent proactively checks invalid moves, UI freezes, numerical inconsistencies, and termination conditions. This approach reveals behavioral faults that unit test-based evaluation would miss.
\end{itemize}

\subsection{Evaluation Metrics}
\label{sec:evaluation_metrics}
Inspired by Pass@k~\cite{chen2021evaluating}, we propose Exec@k and Play@k, which estimate the probability that at least one of the top-\(k\) samples satisfies a task-specific criterion (i.e., successful execution or logically correct gameplay).
For each problem, we draw \(n\) samples and let \(c_{\text{succeed}}\) denote the number of samples meeting the criterion; we then compute the unbiased estimator:
\begin{equation}
\label{equ:success_rate}
\text{[Exec, Pass, Play]@}k := \mathop{\mathbb{E}}\limits_{\text{Problems}} \left[ 1 - \frac{\binom{n-c_{\text{succeed}}}{k}}{\binom{n}{k}} \right]
\end{equation}
We evaluate four dimensions of code generation under this framework.
\textit{Exec@k}: \revise{at least one of the top-\(k\) samples executes successfully without syntax, or compile errors. 
We employ \texttt{Exec@k} and \texttt{Pass@k} following prior works (HumanEval~\cite{chen2021evaluating}, CODERANKER~\cite{inala2022fault}). 
As in Section~\ref{sec:evaluation_workflow}, our evaluation follows the pipeline (\texttt{Exec@k} $\to$ \texttt{Pass@k} $\to$ \texttt{Play@k}). A sample is evaluated for the next stage \textit{only if} it passes the previous one. If the program failed the compilation before running, it is considered failed (Exec@k);
\textit{Pass@k}: at least one of the top-\(k\) samples passes all provided test cases;
\textit{Play@k}: at least one of the top-\(k\) samples exhibits correct GUI behavior under interactive testing by our GUI Behavioral Testing. \texttt{Play@k} is further enforced by constraints (e.g., ``Did the game crash?'') verified by logs and final-state screenshots.}
\revise{
To answer ``Are we maximizing successful results with minimal token usage?'', we propose Efficiency@k, which measures the effectiveness-to-cost ratio:
\begin{equation}
\text{Efficiency@}k := \frac{\text{Play@}k}{AT_k}, \qquad AT_k := \frac{\text{Token}_k}{N \cdot 10^{3}}
\end{equation}
where $AT_k$ denotes the average number of tokens consumed per problem (in thousands), $\text{Token}_k$ is the total token usage over all $N$ problems when generating $k$ samples per problem, and the $10^{3}$ factor keeps the numerator and denominator on comparable scales (matching the caption of Table~\ref{tab:rq3_new}).
For $k=1$, Efficiency@$k$ simplifies through the combinatorial formula in Equation~\ref{equ:success_rate}:
\begin{equation}
\text{Efficiency@1} = \frac{\text{Play@1}}{AT_1} = \frac{\mathbb{E}[c_{\text{succeed}}/n]}{AT_1} = \frac{\text{Succeed}_1 / N}{\text{Token}_1 / N} = \frac{\text{Succeed}_1}{\text{Token}_1},
\end{equation}
where $c_{\text{succeed}}$ is the number of successful samples per problem and $n$ is the number of samples per problem.}

\subsection{Preliminary Study 1: PlayTester Effectiveness}
\label{sec:playtester_validation}

\revise{To establish PlayTester as a reliable evaluation framework, 
we conducted validation through human evaluation and statistical analysis 
before using it to evaluate code generation methods.}
We manually verified 100 successes and 100 failures identified by PlayTester, all randomly selected. 
\revise{To mitigate order bias, these samples were presented to evaluators in a randomized order.} 
We recruited 3 software engineers with >5 years of experience in GUI development to serve as independent evaluators, treating human judgment as the ground truth.
The evaluation reveals a 16\% false-negative rate and a 5\% false-positive rate.
To quantify the consistency between PlayTester and human evaluators, 
we calculated Krippendorff's Alpha ($\alpha=0.790$) and Kendall's Tau-b ($\tau_b=0.795$). 
These statistical measures confirm substantial agreement and establish PlayTester as a reliable evaluation tool for GUI behavioral correctness.
This demonstrates that PlayTester can serve as an automated evaluation framework before being incorporated into any code generation system.

\subsection{Preliminary Study 2: Method Performance}
\label{sec:preliminary}
Having established PlayTester as a reliable evaluation framework in Section~\ref{sec:playtester_validation}, we now employ it to assess the performance of state-of-the-art LLMs and LLM-based methods.

\subsubsection{Experiment Setup}
We evaluate 10 state-of-the-art LLMs and 5 representative LLM-based enhanced methods on \datasetname.
For base LLMs, we use standard few-shot prompting techniques to generate complete, repository-aware code implementations.
For enhanced methods, we evaluate five representative approaches. 
\textit{SCoT} employs structured chain-of-thought reasoning. 
\textit{HCPCoder} uses hierarchical context pruning for repository awareness. 
\textit{MetaGPT} applies multi-agent software development workflows. 
\textit{OpenManus} utilizes collaborative agent coordination. 
\textit{DeepCode} leverages specialized code understanding and generation capabilities.
Each generated solution undergoes evaluation across three progressive criteria using PlayTester as the evaluation framework. 
\textit{Exec@k} measures basic executability without runtime crashes. 
\textit{Pass@k} evaluates correctness against provided unit tests. 
\textit{Play@k} assesses semantic correctness through interactive GUI testing.

\begin{table*}[htbp!]
    \centering
    \caption{\revise{Performance (\%) of LLMs and LLM-based enhanced methods on \datasetname.} \revise{We report mean values over 5 independent runs with 95\% confidence intervals calculated using the Student's $t$-distribution.}}
    \label{tab:rq1}
    \resizebox{\linewidth}{!}
    {
        \begin{tabular}{lc|cccccc|cccccc|cccccc}
        \toprule
        \multirow{2}{*}{\textbf{LLMs} / \textbf{Methods}} & \multirow{2}{*}{\textbf{Size}} & \multicolumn{6}{c|}{\textbf{Python}} & \multicolumn{6}{c|}{\textbf{JavaScript}} & \multicolumn{6}{c}{\textbf{TypeScript}} \\
        \cmidrule(lr){3-8} \cmidrule(lr){9-14} \cmidrule(lr){15-20}
        & & \textbf{Exec@1} & \textbf{Exec@3} & \textbf{Pass@1} & \textbf{Pass@3} & \textbf{Play@1} & \textbf{Play@3} & \textbf{Exec@1} & \textbf{Exec@3} & \textbf{Pass@1} & \textbf{Pass@3} & \textbf{Play@1} & \textbf{Play@3} & \textbf{Exec@1} & \textbf{Exec@3} & \textbf{Pass@1} & \textbf{Pass@3} & \textbf{Play@1} & \textbf{Play@3} \\
        \midrule
        \multicolumn{20}{l}{\textit{\textbf{Base LLMs}}} \\
        \midrule
        \cellcolor[HTML]{e1ecff}{GPT-4o-mini} & \cellcolor[HTML]{e1ecff}{-} & \cellcolor[HTML]{e1ecff}{10.3 $\pm$ 1.1} & \cellcolor[HTML]{e1ecff}{12.7 $\pm$ 1.4} & \cellcolor[HTML]{e1ecff}{2.9 $\pm$ 0.3} & \cellcolor[HTML]{e1ecff}{5.2 $\pm$ 0.5} & \cellcolor[HTML]{e1ecff}{2.1 $\pm$ 0.2} & \cellcolor[HTML]{e1ecff}{2.6 $\pm$ 0.4} & \cellcolor[HTML]{e1ecff}{11.1 $\pm$ 1.3} & \cellcolor[HTML]{e1ecff}{13.4 $\pm$ 1.2} & \cellcolor[HTML]{e1ecff}{3.1 $\pm$ 0.4} & \cellcolor[HTML]{e1ecff}{5.5 $\pm$ 0.6} & \cellcolor[HTML]{e1ecff}{2.3 $\pm$ 0.3} & \cellcolor[HTML]{e1ecff}{2.9 $\pm$ 0.5} & \cellcolor[HTML]{e1ecff}{8.7 $\pm$ 0.9} & \cellcolor[HTML]{e1ecff}{10.2 $\pm$ 1.1} & \cellcolor[HTML]{e1ecff}{2.2 $\pm$ 0.2} & \cellcolor[HTML]{e1ecff}{3.9 $\pm$ 0.4} & \cellcolor[HTML]{e1ecff}{1.6 $\pm$ 0.2} & \cellcolor[HTML]{e1ecff}{2.0 $\pm$ 0.3} \\
        GPT-5-mini & - & 12.4 $\pm$ 2.0 & 13.7 $\pm$ 1.6 & 6.4 $\pm$ 0.3 & 6.7 $\pm$ 0.5 & 4.3 $\pm$ 0.5 & 5.2 $\pm$ 0.6 & 13.2 $\pm$ 1.8 & 14.5 $\pm$ 1.7 & 6.7 $\pm$ 0.4 & 7.1 $\pm$ 0.6 & 4.5 $\pm$ 0.6 & 5.5 $\pm$ 0.7 & 10.3 $\pm$ 1.6 & 11.4 $\pm$ 1.3 & 5.1 $\pm$ 0.3 & 5.4 $\pm$ 0.4 & 3.4 $\pm$ 0.4 & 4.1 $\pm$ 0.5 \\
        \cellcolor[HTML]{e1ecff}{Grok-3-mini} & \cellcolor[HTML]{e1ecff}{-} & \cellcolor[HTML]{e1ecff}{13.9 $\pm$ 2.0} & \cellcolor[HTML]{e1ecff}{16.6 $\pm$ 1.8} & \cellcolor[HTML]{e1ecff}{7.0 $\pm$ 0.6} & \cellcolor[HTML]{e1ecff}{8.1 $\pm$ 0.9} & \cellcolor[HTML]{e1ecff}{4.6 $\pm$ 1.0} & \cellcolor[HTML]{e1ecff}{5.8 $\pm$ 0.6} & \cellcolor[HTML]{e1ecff}{14.7 $\pm$ 2.1} & \cellcolor[HTML]{e1ecff}{17.3 $\pm$ 1.9} & \cellcolor[HTML]{e1ecff}{7.4 $\pm$ 0.7} & \cellcolor[HTML]{e1ecff}{8.5 $\pm$ 1.0} & \cellcolor[HTML]{e1ecff}{4.8 $\pm$ 1.1} & \cellcolor[HTML]{e1ecff}{6.1 $\pm$ 0.7} & \cellcolor[HTML]{e1ecff}{11.5 $\pm$ 1.7} & \cellcolor[HTML]{e1ecff}{13.8 $\pm$ 1.5} & \cellcolor[HTML]{e1ecff}{5.6 $\pm$ 0.5} & \cellcolor[HTML]{e1ecff}{6.5 $\pm$ 0.7} & \cellcolor[HTML]{e1ecff}{3.7 $\pm$ 0.8} & \cellcolor[HTML]{e1ecff}{4.6 $\pm$ 0.5} \\
        Claude-Sonnet-3.7 & - & 10.8 $\pm$ 1.7 & 13.1 $\pm$ 1.1 & 6.1 $\pm$ 1.1 & 9.6 $\pm$ 1.8 & 4.7 $\pm$ 0.4 & 7.5 $\pm$ 1.0 & 11.5 $\pm$ 1.8 & 13.9 $\pm$ 1.2 & 6.5 $\pm$ 1.2 & 10.1 $\pm$ 1.9 & 5.0 $\pm$ 0.5 & 7.9 $\pm$ 1.1 & 9.0 $\pm$ 1.4 & 10.9 $\pm$ 0.9 & 4.9 $\pm$ 0.9 & 7.7 $\pm$ 1.4 & 3.8 $\pm$ 0.3 & 6.0 $\pm$ 0.8 \\
        \cellcolor[HTML]{e1ecff}{Claude-Sonnet-4} & \cellcolor[HTML]{e1ecff}{-} & \cellcolor[HTML]{e1ecff}{\textbf{17.9 $\pm$ 2.1}} & \cellcolor[HTML]{e1ecff}{\textbf{18.6 $\pm$ 3.5}} & \cellcolor[HTML]{e1ecff}{\textbf{10.1 $\pm$ 0.7}} & \cellcolor[HTML]{e1ecff}{\textbf{13.0 $\pm$ 2.2}} & \cellcolor[HTML]{e1ecff}{6.4 $\pm$ 0.6} & \cellcolor[HTML]{e1ecff}\textbf{{9.9 $\pm$ 0.8}} & \cellcolor[HTML]{e1ecff}{\textbf{19.3 $\pm$ 2.3}} & \cellcolor[HTML]{e1ecff}{19.7 $\pm$ 3.7} & \cellcolor[HTML]{e1ecff}{\textbf{10.6 $\pm$ 0.8}} & \cellcolor[HTML]{e1ecff}{\textbf{13.7 $\pm$ 2.3}} & \cellcolor[HTML]{e1ecff}{6.7 $\pm$ 0.7} & \cellcolor[HTML]{e1ecff}\textbf{{10.4 $\pm$ 0.9}} & \cellcolor[HTML]{e1ecff}{\textbf{14.5 $\pm$ 1.7}} & \cellcolor[HTML]{e1ecff}{\textbf{16.7 $\pm$ 2.9}} & \cellcolor[HTML]{e1ecff}{\textbf{8.1 $\pm$ 0.6}} & \cellcolor[HTML]{e1ecff}{\textbf{10.4 $\pm$ 1.8}} & \cellcolor[HTML]{e1ecff}{5.1 $\pm$ 0.5} & \cellcolor[HTML]{e1ecff}{\textbf{7.9 $\pm$ 0.6}} \\
        GPT-4o & - & 13.7 $\pm$ 3.4 & 13.8 $\pm$ 1.6 & 8.1 $\pm$ 0.5 & 8.9 $\pm$ 0.8 & 3.8 $\pm$ 0.4 & 6.7 $\pm$ 1.1 & 14.6 $\pm$ 3.2 & 14.6 $\pm$ 1.7 & 8.5 $\pm$ 0.6 & 9.4 $\pm$ 0.9 & 4.0 $\pm$ 0.5 & 7.1 $\pm$ 1.2 & 11.3 $\pm$ 2.8 & 11.5 $\pm$ 1.3 & 6.5 $\pm$ 0.4 & 7.1 $\pm$ 0.6 & 3.0 $\pm$ 0.3 & 5.4 $\pm$ 0.9 \\
        \cellcolor[HTML]{e1ecff}{GPT-5} & \cellcolor[HTML]{e1ecff}{-} & \cellcolor[HTML]{e1ecff}{17.4 $\pm$ 1.9} & \cellcolor[HTML]{e1ecff}{17.5 $\pm$ 2.2} & \cellcolor[HTML]{e1ecff}{8.6 $\pm$ 1.1} & \cellcolor[HTML]{e1ecff}{10.2 $\pm$ 0.8} & \cellcolor[HTML]{e1ecff}{\textbf{6.6 $\pm$ 1.6}} & \cellcolor[HTML]{e1ecff}{6.9 $\pm$ 1.5} & \cellcolor[HTML]{e1ecff}{18.5 $\pm$ 2.0} & \cellcolor[HTML]{e1ecff}{18.6 $\pm$ 2.3} & \cellcolor[HTML]{e1ecff}{9.1 $\pm$ 1.2} & \cellcolor[HTML]{e1ecff}{10.7 $\pm$ 0.9} & \cellcolor[HTML]{e1ecff}{\textbf{8.0 $\pm$ 1.7}} & \cellcolor[HTML]{e1ecff}{8.9 $\pm$ 1.6} & \cellcolor[HTML]{e1ecff}{14.4 $\pm$ 1.6} & \cellcolor[HTML]{e1ecff}{14.5 $\pm$ 1.8} & \cellcolor[HTML]{e1ecff}{6.9 $\pm$ 0.9} & \cellcolor[HTML]{e1ecff}{8.2 $\pm$ 0.6} & \cellcolor[HTML]{e1ecff}{\textbf{6.1 $\pm$ 1.3}} & \cellcolor[HTML]{e1ecff}{5.2 $\pm$ 1.2} \\
        Qwen3-Coder & 480B & 14.0 $\pm$ 1.8 & \textbf{18.8 $\pm$ 4.7} & 5.2 $\pm$ 0.8 & 5.9 $\pm$ 1.1 & 4.9 $\pm$ 0.8 & 6.1 $\pm$ 0.4 & 14.8 $\pm$ 1.9 & \textbf{19.8 $\pm$ 4.9} & 5.5 $\pm$ 0.9 & 6.2 $\pm$ 1.2 & 5.2 $\pm$ 0.9 & 6.4 $\pm$ 0.5 & 11.6 $\pm$ 1.5 & 15.6 $\pm$ 3.9 & 4.2 $\pm$ 0.6 & 4.7 $\pm$ 0.9 & 3.9 $\pm$ 0.6 & 4.9 $\pm$ 0.3 \\
        \cellcolor[HTML]{e1ecff}{GLM-4.5} & \cellcolor[HTML]{e1ecff}{355B} & \cellcolor[HTML]{e1ecff}{7.6 $\pm$ 2.1} & \cellcolor[HTML]{e1ecff}{\textbf{17.8 $\pm$ 0.9}} & \cellcolor[HTML]{e1ecff}{7.3 $\pm$ 1.1} & \cellcolor[HTML]{e1ecff}{7.1 $\pm$ 0.9} & \cellcolor[HTML]{e1ecff}{5.9 $\pm$ 1.7} & \cellcolor[HTML]{e1ecff}{6.3 $\pm$ 0.7} & \cellcolor[HTML]{e1ecff}{8.1 $\pm$ 2.2} & \cellcolor[HTML]{e1ecff}{18.7 $\pm$ 1.0} & \cellcolor[HTML]{e1ecff}{7.7 $\pm$ 1.2} & \cellcolor[HTML]{e1ecff}{8.0 $\pm$ 1.0} & \cellcolor[HTML]{e1ecff}{6.2 $\pm$ 1.8} & \cellcolor[HTML]{e1ecff}{6.6 $\pm$ 0.8} & \cellcolor[HTML]{e1ecff}{6.3 $\pm$ 1.7} & \cellcolor[HTML]{e1ecff}{14.7 $\pm$ 0.7} & \cellcolor[HTML]{e1ecff}{5.8 $\pm$ 0.9} & \cellcolor[HTML]{e1ecff}{6.1 $\pm$ 0.7} & \cellcolor[HTML]{e1ecff}{4.7 $\pm$ 1.4} & \cellcolor[HTML]{e1ecff}{5.0 $\pm$ 0.6} \\
        DeepSeek-V3 & 671B & 11.7 $\pm$ 2.2 & 15.1 $\pm$ 1.1 & 5.8 $\pm$ 0.9 & 7.1 $\pm$ 0.5 & 5.0 $\pm$ 0.3 & 7.2 $\pm$ 1.1 & 12.5 $\pm$ 2.3 & 16.0 $\pm$ 1.2 & 6.1 $\pm$ 1.0 & 7.5 $\pm$ 0.6 & 5.3 $\pm$ 0.4 & 7.6 $\pm$ 1.2 & 9.7 $\pm$ 1.8 & 12.5 $\pm$ 0.9 & 4.6 $\pm$ 0.7 & 5.7 $\pm$ 0.4 & 4.0 $\pm$ 0.2 & 5.8 $\pm$ 0.9 \\
        \midrule
        \multicolumn{20}{l}{\textit{\textbf{LLM-based Enhanced Methods (GPT-5-mini as backbone LLM)}}} \\
        \midrule
        \cellcolor[HTML]{e1ecff}{SCoT~\cite{li2025structured}} & \cellcolor[HTML]{e1ecff}{-} & \cellcolor[HTML]{e1ecff}{13.8 $\pm$ 1.3} & \cellcolor[HTML]{e1ecff}{15.2 $\pm$ 1.3} & \cellcolor[HTML]{e1ecff}{4.7 $\pm$ 0.5} & \cellcolor[HTML]{e1ecff}{7.0 $\pm$ 0.6} & \cellcolor[HTML]{e1ecff}{4.8 $\pm$ 0.3} & \cellcolor[HTML]{e1ecff}{6.0 $\pm$ 0.5} & \cellcolor[HTML]{e1ecff}{14.6 $\pm$ 1.5} & \cellcolor[HTML]{e1ecff}{16.2 $\pm$ 1.2} & \cellcolor[HTML]{e1ecff}{5.9 $\pm$ 0.7} & \cellcolor[HTML]{e1ecff}{7.6 $\pm$ 1.2} & \cellcolor[HTML]{e1ecff}{5.2 $\pm$ 0.6} & \cellcolor[HTML]{e1ecff}{6.1 $\pm$ 1.0} & \cellcolor[HTML]{e1ecff}{11.3 $\pm$ 1.2} & \cellcolor[HTML]{e1ecff}{12.7 $\pm$ 0.9} & \cellcolor[HTML]{e1ecff}{4.5 $\pm$ 0.5} & \cellcolor[HTML]{e1ecff}{5.8 $\pm$ 0.9} & \cellcolor[HTML]{e1ecff}{3.9 $\pm$ 0.4} & \cellcolor[HTML]{e1ecff}{4.6 $\pm$ 0.7} \\
        HCPCoder~\cite{zhang2025hierarchical} & - & 12.3 $\pm$ 2.1 & 12.8 $\pm$ 2.9 & 1.7 $\pm$ 0.1 & 3.5 $\pm$ 0.5 & 0.3 $\pm$ 0.0 & 0.3 $\pm$ 0.1 & 14.1 $\pm$ 0.6 & 14.6 $\pm$ 3.1 & 1.7 $\pm$ 0.5 & 3.5 $\pm$ 0.3 & 0.3 $\pm$ 0.1 & 0.3 $\pm$ 0.1 & 10.9 $\pm$ 0.4 & 11.4 $\pm$ 2.5 & 1.3 $\pm$ 0.3 & 2.6 $\pm$ 0.2 & 0.2 $\pm$ 0.1 & 0.2 $\pm$ 0.1 \\
        \cellcolor[HTML]{e1ecff}{MetaGPT~\cite{hong2024metagpt}} & \cellcolor[HTML]{e1ecff}{-} & \cellcolor[HTML]{e1ecff}{12.6 $\pm$ 0.7} & \cellcolor[HTML]{e1ecff}{13.0 $\pm$ 1.8} & \cellcolor[HTML]{e1ecff}{6.6 $\pm$ 0.5} & \cellcolor[HTML]{e1ecff}{10.3 $\pm$ 1.1} & \cellcolor[HTML]{e1ecff}{4.0 $\pm$ 0.7} & \cellcolor[HTML]{e1ecff}{4.4 $\pm$ 0.6} & \cellcolor[HTML]{e1ecff}{13.3 $\pm$ 1.9} & \cellcolor[HTML]{e1ecff}{13.6 $\pm$ 2.5} & \cellcolor[HTML]{e1ecff}{7.5 $\pm$ 1.9} & \cellcolor[HTML]{e1ecff}{9.9 $\pm$ 1.6} & \cellcolor[HTML]{e1ecff}{4.5 $\pm$ 0.7} & \cellcolor[HTML]{e1ecff}{4.9 $\pm$ 0.5} & \cellcolor[HTML]{e1ecff}{10.3 $\pm$ 1.5} & \cellcolor[HTML]{e1ecff}{10.6 $\pm$ 2.0} & \cellcolor[HTML]{e1ecff}{5.7 $\pm$ 1.4} & \cellcolor[HTML]{e1ecff}{7.5 $\pm$ 1.2} & \cellcolor[HTML]{e1ecff}{3.4 $\pm$ 0.5} & \cellcolor[HTML]{e1ecff}{3.7 $\pm$ 0.3} \\
        OpenManus~\cite{openmanus2025} & - & 12.3 $\pm$ 0.9 & 15.4 $\pm$ 2.6 & 8.1 $\pm$ 2.0 & 12.2 $\pm$ 1.2 & 5.3 $\pm$ 1.2 & 5.6 $\pm$ 0.9 & 13.1 $\pm$ 1.0 & 16.3 $\pm$ 2.7 & 8.5 $\pm$ 2.1 & 12.8 $\pm$ 1.3 & 5.6 $\pm$ 1.3 & 5.9 $\pm$ 1.0 & 10.2 $\pm$ 0.7 & 12.8 $\pm$ 2.2 & 6.5 $\pm$ 1.6 & 9.8 $\pm$ 1.0 & 4.2 $\pm$ 1.0 & 4.5 $\pm$ 0.7 \\
        \cellcolor[HTML]{e1ecff}{DeepCode~\cite{deepcode2025}} & \cellcolor[HTML]{e1ecff}{-} & \cellcolor[HTML]{e1ecff}{\textbf{17.1 $\pm$ 3.8}} & \cellcolor[HTML]{e1ecff}{\textbf{17.9 $\pm$ 6.4}} & \cellcolor[HTML]{e1ecff}{\textbf{10.5 $\pm$ 1.3}} & \cellcolor[HTML]{e1ecff}{\textbf{14.2 $\pm$ 2.2}} & \cellcolor[HTML]{e1ecff}{\textbf{6.0 $\pm$ 1.0}} & \cellcolor[HTML]{e1ecff}{\textbf{6.4 $\pm$ 1.0}} & \cellcolor[HTML]{e1ecff}{\textbf{18.2 $\pm$ 4.0}} & \cellcolor[HTML]{e1ecff}{\textbf{19.0 $\pm$ 6.7}} & \cellcolor[HTML]{e1ecff}{\textbf{11.0 $\pm$ 1.4}} & \cellcolor[HTML]{e1ecff}{\textbf{14.9 $\pm$ 2.3}} & \cellcolor[HTML]{e1ecff}{\textbf{5.9 $\pm$ 1.6}} & \cellcolor[HTML]{e1ecff}{\textbf{6.7 $\pm$ 1.1}} & \cellcolor[HTML]{e1ecff}{\textbf{14.1 $\pm$ 3.1}} & \cellcolor[HTML]{e1ecff}{\textbf{14.8 $\pm$ 5.3}} & \cellcolor[HTML]{e1ecff}{\textbf{8.4 $\pm$ 1.0}} & \cellcolor[HTML]{e1ecff}{\textbf{11.4 $\pm$ 1.8}} & \cellcolor[HTML]{e1ecff}{\textbf{5.3 $\pm$ 0.8}} & \cellcolor[HTML]{e1ecff}{\textbf{6.0 $\pm$ 0.4}} \\
        \bottomrule
        \end{tabular}
    }
\end{table*}

\subsubsection{Results}
Table~\ref{tab:rq1} presents the performance of 10 state-of-the-art LLMs and 5 representative LLM-based enhanced methods on \datasetname across three programming languages and all evaluation metrics.
The results reveal striking performance degradation as evaluation criteria become more stringent across all baselines, with notable cross-language performance variations.

\noindent\textit{\underline{Base LLM Performance:}} 
Among base LLMs, Claude-Sonnet-4 demonstrates the strongest performance across all three languages, maintaining its lead in both execution and behavioral validation metrics.
However, even top-performing models achieve relatively modest behavioral correctness rates, with the best Play@3 scores remaining in the single digits for TypeScript implementations.
Most models exhibit significant performance drops from execution to behavioral validation across all languages.
Notably, the execution-to-behavior gap widens progressively from JavaScript to Python to TypeScript, suggesting that syntactic correctness does not reliably predict behavioral validity in statically-typed GUI applications.

\noindent\textit{\underline{LLM-based Enhanced Methods:}} 
Evaluation of state-of-the-art LLM-based enhancement methods reveals limited and inconsistent improvements over base models across all three languages.
Prompting-based methods (e.g., SCoT) show marginal improvements in Python and JavaScript but fail to bridge the performance gap for TypeScript implementations.
Repository-aware approaches (e.g., HCPCoder) demonstrate catastrophic results across all languages, particularly in behavioral validation, indicating that context retrieval alone cannot address the semantic complexity of GUI applications.
Multi-agent frameworks exhibit mixed results, with some methods (OpenManus) showing modest improvements in Python and JavaScript while others (MetaGPT) experience performance degradation relative to base models.
DeepCode, despite specialized code understanding capabilities, achieves strong execution success but demonstrates limited behavioral validation improvements, with this pattern consistent across all three languages.
Critically, no enhanced method successfully narrows the cross-language performance gap, suggesting that current enhancement strategies do not adequately address language-specific challenges in GUI code generation.

\intuition{
\textbf{Answer to Preliminary Study:} Evaluation across state-of-the-art LLMs and methods reveals their limitations on \datasetname. The consistent failures across existing approaches indicate that repository-level GUI code generation remains challenging, even with prompting and agentic approaches.
}

\section{Our Approach: \mytitle}
\begin{figure}[htbp]
    \centering
    \includegraphics[width=0.88\textwidth]{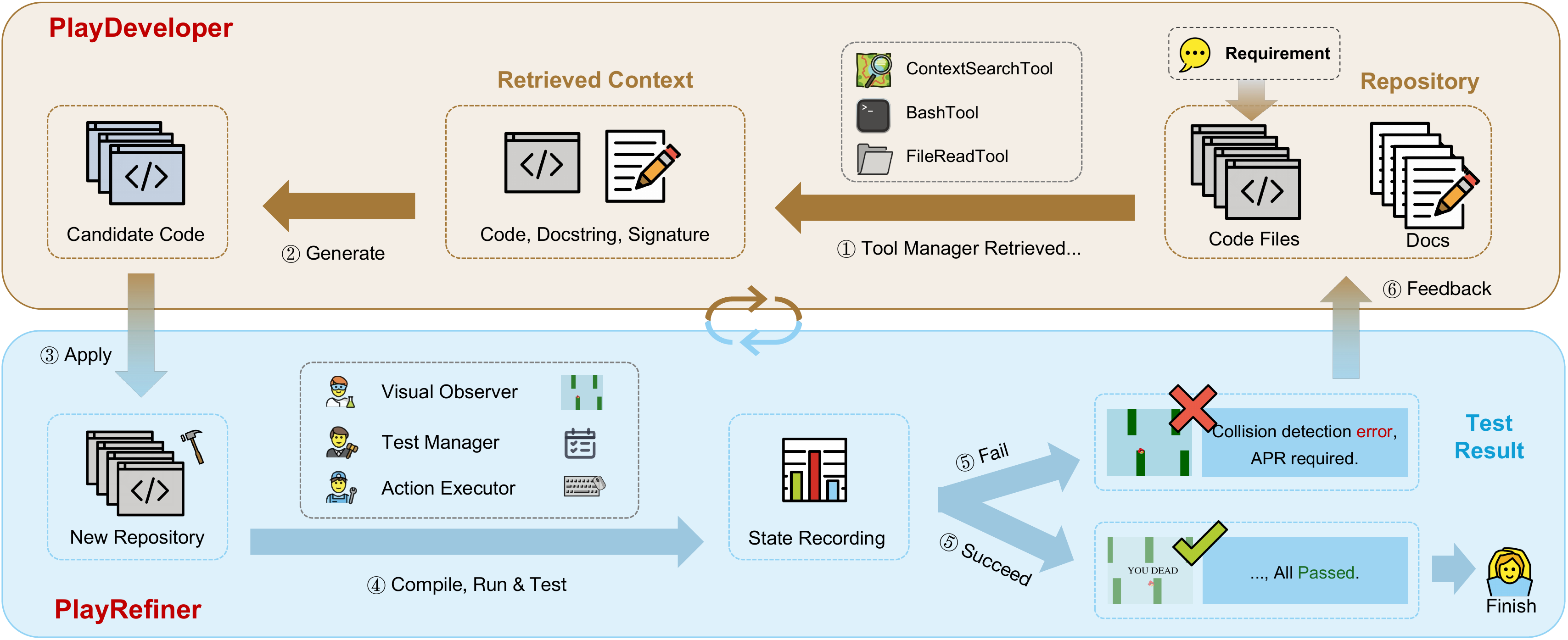}
    \caption{The overview of \mytitle.}
    \label{fig:approach}
\end{figure}

To address the critical challenge of repository-aware GUI application code generation,
we propose \mytitle, a novel multi-agent framework that leverages two specialized agents as shown in Fig.~\ref{fig:approach}: 
(1) PlayDeveloper: a repository-aware agent for code generation, and
(2) PlayRefiner: an automated program repair (APR) agent to iteratively refine code based on behavioral testing feedback.

\subsection{Multi-Agent Collaboration Workflow}
\label{sec:multi-agent_workflow}
\revise{We detail the collaboration workflow between the two agents in \mytitle{} 
and then summarize the shared processes and telemetry that support this workflow.}

\subsubsection{Workflow Phases} 
\revise{The two agents collaborate through a structured test\,\&\,repair cycle to achieve 
both syntactic correctness and behavioral alignment:}

\begin{enumerate}[leftmargin=*]
    \item \textbf{Context-Aware Generation.} 
    PlayDeveloper receives specifications and generates repository-aware GUI application code using retrieved patterns and module structures.
    \item \revise{\textbf{Behavioral Testing.}} 
    \revise{The generated application is evaluated through automated behavioral testing (Section~\ref{sec:playtester}): 
    the Visual Observer Module captures application state, 
    the Test Manager plans interaction sequences using vision-language analysis, 
    and the Action Executor Module executes tests to collect behavioral signals.}
    \item \textbf{Diagnosis \& Repair.} 
    \revise{PlayRefiner analyzes execution traces and testing feedback from the behavioral testing modules, 
    synthesizes patches with repository context, and applies fixes with compilation/runtime checks.}
    \item \textbf{Iterative Feedback.} 
    \revise{The updated application is re-evaluated through automated behavioral testing, 
    checking behavior against specifications, 
    including interactive semantics 
    (e.g., collision handling, event responses, and state transitions).}
    Each cycle refines code quality and testing strategy based on accumulated feedback.
    This process continues until the specified behavioral criteria are met or the iteration limit is reached. \revise{Concretely, the generate-refine loop terminates early once the application passes all per-sample behavioral checks enforced by PlayTester, and otherwise runs up to a maximum of $T=6$ iterations.} 
\end{enumerate}

\subsubsection{Trajectory Recording} 
The multi-agent framework maintains comprehensive execution logs through the \texttt{AgentTrajectory} tool.
This tool logs LLM interactions, tool usage, token consumption, screenshots, actions, and decision points. 
This comprehensive logging enables diagnosis, ablation studies, reproducibility, and APR prioritization.

\subsubsection{Sandboxing \& Determinism}
Applications execute in sandboxed environments with deterministic seeding and controlled timing. 
The framework emits standardized logs to ensure fair, reproducible comparisons and to provide precise failure signals to 
the Testing and APR agents.

\subsection{PlayDeveloper}

PlayDeveloper implements a tool-based architecture dedicated to repository-aware code generation. 
The agent supports multiple LLM providers (e.g., OpenAI, Anthropic) 
and employs a modular tool ecosystem for context collection and code generation. 
The key components of \texttt{ToolManager} include:
(1) \texttt{ContextSearchTool} for retrieving relevant code examples 
and import patterns from the repository context using grep-based searching, 
(2) \texttt{FileReadTool} for file operations, 
(3) \texttt{BashTool} for executing shell commands, 
and (4) \texttt{ConversationTool} for maintaining dialogue sessions. \revise{We use few-shot prompting with standard requirement-code examples.}

\subsection{PlayRefiner}
This agent performs APR driven by behavioral feedback and execution traces. 
The agent coordinates validation tooling and orchestrates iterative fixes until behavioral criteria are satisfied.

\subsubsection{Repair Tooling}
PlayRefiner coordinates three core tools: 
(1) \texttt{ContextSearcher} for retrieving repository-aware APIs and import patterns during code repair, 
(2) \texttt{Validator} for syntax/AST checks and fast compile gating, and 
(3) \texttt{Executor} for executing the target program in a sandbox to capture runtime and behavioral signals.

\subsubsection{Repair Workflow}
The APR loop proceeds in five phases: 
(1)~\emph{Diagnosis} aggregates compiler output, runtime logs, 
\revise{and behavioral testing reports 
(screenshots, actions, and unexpected behaviors)} into actionable failure summaries, 
(2)~\emph{Patch Generation} proposes minimal edits guided by retrieved context, 
(3)~\emph{Patch Application} writes changes atomically to the repository, 
(4)~\revise{\emph{Build \& Runtime Validation} compiles and executes the application, followed by behavioral re-evaluation through the testing modules (Section~\ref{sec:playtester}), and}
(5)~\emph{Iterative Refinement} repeats up to a fixed budget or until behavioral criteria are met.

\section{Results}

To evaluate the effectiveness and efficiency of \mytitle, we formulate three research questions:

\begin{itemize}[leftmargin=*]
\item \textbf{RQ-1: Effectiveness Study.} \textit{How does the performance of \mytitle compare with the baselines?}
\item \textbf{RQ-2: Efficiency Study.} \textit{How does the efficiency of \mytitle compare with the baselines?}
\item \textbf{RQ-3: Ablation Study.} \textit{How do different components and models affect \mytitle's effectiveness?}
\end{itemize}

\revise{\noindent\textbf{Statistical Reporting.}
We report the mean of each metric over 5 independent runs.
Each run uses a distinct predefined random seed assigned to a unique integer, while keeping the LLM sampling temperature fixed at $T = 0.3$, prompts identical without dynamic elements, and all external tools and database states reset before each run.
We quantify uncertainty using 95\% confidence intervals for the mean, computed with the Student's $t$-distribution, which is suitable for small sample sizes.
We observe that many improvements are stable with relatively small intervals, while some close comparisons show overlapping intervals and should be interpreted as comparable rather than decisively better.
The variance mainly comes from stochastic LLM sampling and multi-agent interaction trajectories, and it does not change the overall ranking trends reported in this section.
}

\subsection{RQ-1: How does \mytitle perform compared to baselines?}

\noindent
\textbf{Objective.}
This research question evaluates the effectiveness of \mytitle{} against state-of-the-art repository-aware code generation baselines.
We investigate whether our multi-agent approach, which combines automated program repair with dynamic GUI testing, provides substantial improvements over existing methods that focus solely on enhanced prompting strategies.
Specifically, we compare \mytitle against five representative baselines: 
SCoT~\cite{li2025structured}, 
HCPCoder~\cite{zhang2025hierarchical}, 
MetaGPT~\cite{hong2024metagpt}, 
OpenManus~\cite{openmanus2025}, and DeepCode~\cite{deepcode2025}
evaluating performance across Exec@k, Pass@k, and Play@k to assess both syntactic correctness and behavioral validity.

\noindent
\textbf{Experimental Design.}
We conduct experiments using three different LLMs to ensure fair comparison.
For SCoT, we implement structured chain-of-thought prompting following an 8-step construction pipeline that guides LLMs through systematic reasoning before code generation. 
The approach uses fixed demonstration examples covering sequential, branch, and loop programming structures, embedding step-by-step reasoning as line comments within the prompt template.
The method first provides high-level instructions, then presents demonstration examples with structured reasoning patterns, and finally requests the model to follow the same reasoning approach for the target task.
HCPCoder implements hierarchical context pruning with repository-aware prompt generation.
Both SCoT and HCPCoder operate as single-shot generation methods without iterative refinement capabilities.
The five baselines are evaluated across three backbone LLMs.

\noindent
\textbf{Results.}
\revise{Table~\ref{tab:rq2} presents cross-language performance across three LLMs (GPT-5-mini, Claude-Sonnet-3.7, Qwen3-Coder) evaluating Python, JavaScript, and TypeScript.
JavaScript demonstrates comparable or superior performance to Python across most methods (e.g., \mytitle: 30.9\% vs 26.8\% Exec@3), while TypeScript exhibits systematically lower performance, since there are complex MMORPG games in our benchmark implemented in TypeScript.
With GPT-5-mini, \mytitle substantially outperforms all baselines across languages: compared to SCoT, improvements include 11.6pp in Exec@3 and 12.7pp in Pass@3; compared to HCPCoder, 14.0pp and 16.2pp respectively.
Notably, all baselines struggle with behavioral validation (Play@k).
HCPCoder achieves near-zero Play@k across all languages despite sophisticated repository retrieval and token consumption.
Prompting methods (SCoT) and multi-agent frameworks (MetaGPT, OpenManus) show limited improvements, failing to capture GUI requirements regardless of language.
Cross-language patterns remain stable across LLM backbones, confirming language-specific trends reflect code generation difficulty rather than model biases.
To address evaluator bias concerns~\cite{zheng2023judging}, we validated with Claude-Sonnet-3.7, observing minimal ranking impact (Play@3 variations 0.9\%-3.7\%).}

\begin{table*}[htbp!]
    \centering
    \caption{\revise{Performance (\%) of \mytitle compared to baselines across different LLMs. We report mean values over 5 independent runs with 95\% confidence intervals calculated using the Student's $t$-distribution.}
    }
    \resizebox{\linewidth}{!}
    {
        \begin{tabular}{ll|cccccc|cccccc|cccccc}
        \toprule
        \multirow{2}{*}{\textbf{Methods}} & \multirow{2}{*}{\textbf{LLMs}} & \multicolumn{6}{c|}{\textbf{Python}} & \multicolumn{6}{c|}{\textbf{JavaScript}} & \multicolumn{6}{c}{\textbf{TypeScript}} \\
        \cmidrule(lr){3-8} \cmidrule(lr){9-14} \cmidrule(lr){15-20}
        & & \textbf{Exec@1} & \textbf{Exec@3} & \textbf{Pass@1} & \textbf{Pass@3} & \textbf{Play@1} & \textbf{Play@3} & \textbf{Exec@1} & \textbf{Exec@3} & \textbf{Pass@1} & \textbf{Pass@3} & \textbf{Play@1} & \textbf{Play@3} & \textbf{Exec@1} & \textbf{Exec@3} & \textbf{Pass@1} & \textbf{Pass@3} & \textbf{Play@1} & \textbf{Play@3} \\
        \midrule
        \cellcolor[HTML]{e1ecff}{HCPCoder~\cite{zhang2025hierarchical}} & \cellcolor[HTML]{e1ecff}{GPT-5-mini} & \cellcolor[HTML]{e1ecff}{12.3 $\pm$ 2.1} & \cellcolor[HTML]{e1ecff}{12.8 $\pm$ 2.9} & \cellcolor[HTML]{e1ecff}{1.7 $\pm$ 0.1} & \cellcolor[HTML]{e1ecff}{3.5 $\pm$ 0.5} & \cellcolor[HTML]{e1ecff}{0.3 $\pm$ 0.0} & \cellcolor[HTML]{e1ecff}{0.3 $\pm$ 0.1} & \cellcolor[HTML]{e1ecff}{14.1 $\pm$ 0.6} & \cellcolor[HTML]{e1ecff}{14.6 $\pm$ 3.1} & \cellcolor[HTML]{e1ecff}{1.7 $\pm$ 0.5} & \cellcolor[HTML]{e1ecff}{3.5 $\pm$ 0.3} & \cellcolor[HTML]{e1ecff}{0.3 $\pm$ 0.1} & \cellcolor[HTML]{e1ecff}{0.3 $\pm$ 0.1} & \cellcolor[HTML]{e1ecff}{10.9 $\pm$ 0.4} & \cellcolor[HTML]{e1ecff}{11.4 $\pm$ 2.5} & \cellcolor[HTML]{e1ecff}{1.3 $\pm$ 0.3} & \cellcolor[HTML]{e1ecff}{2.6 $\pm$ 0.2} & \cellcolor[HTML]{e1ecff}{0.2 $\pm$ 0.1} & \cellcolor[HTML]{e1ecff}{0.2 $\pm$ 0.1} \\
        \cellcolor[HTML]{e1ecff}{} & \cellcolor[HTML]{e1ecff}{Claude-Sonnet-3.7} & \cellcolor[HTML]{e1ecff}{16.1 $\pm$ 2.6} & \cellcolor[HTML]{e1ecff}{18.3 $\pm$ 2.2} & \cellcolor[HTML]{e1ecff}{4.9 $\pm$ 0.7} & \cellcolor[HTML]{e1ecff}{7.0 $\pm$ 1.4} & \cellcolor[HTML]{e1ecff}{5.0 $\pm$ 0.5} & \cellcolor[HTML]{e1ecff}{6.0 $\pm$ 0.7} & \cellcolor[HTML]{e1ecff}{18.5 $\pm$ 2.9} & \cellcolor[HTML]{e1ecff}{21.3 $\pm$ 2.6} & \cellcolor[HTML]{e1ecff}{5.8 $\pm$ 1.0} & \cellcolor[HTML]{e1ecff}{8.3 $\pm$ 1.6} & \cellcolor[HTML]{e1ecff}{5.4 $\pm$ 0.7} & \cellcolor[HTML]{e1ecff}{6.5 $\pm$ 0.9} & \cellcolor[HTML]{e1ecff}{13.2 $\pm$ 2.1} & \cellcolor[HTML]{e1ecff}{15.0 $\pm$ 1.8} & \cellcolor[HTML]{e1ecff}{3.9 $\pm$ 0.6} & \cellcolor[HTML]{e1ecff}{5.6 $\pm$ 1.1} & \cellcolor[HTML]{e1ecff}{4.0 $\pm$ 0.4} & \cellcolor[HTML]{e1ecff}{4.8 $\pm$ 0.6} \\
        \cellcolor[HTML]{e1ecff}{} & \cellcolor[HTML]{e1ecff}{Qwen3-Coder} & \cellcolor[HTML]{e1ecff}{15.0 $\pm$ 1.9} & \cellcolor[HTML]{e1ecff}{18.1 $\pm$ 3.2} & \cellcolor[HTML]{e1ecff}{6.7 $\pm$ 1.2} & \cellcolor[HTML]{e1ecff}{12.0 $\pm$ 2.0} & \cellcolor[HTML]{e1ecff}{4.9 $\pm$ 1.0} & \cellcolor[HTML]{e1ecff}{6.5 $\pm$ 1.8} & \cellcolor[HTML]{e1ecff}{17.3 $\pm$ 2.3} & \cellcolor[HTML]{e1ecff}{20.7 $\pm$ 3.6} & \cellcolor[HTML]{e1ecff}{7.9 $\pm$ 1.5} & \cellcolor[HTML]{e1ecff}{13.8 $\pm$ 2.4} & \cellcolor[HTML]{e1ecff}{5.6 $\pm$ 1.2} & \cellcolor[HTML]{e1ecff}{7.4 $\pm$ 2.0} & \cellcolor[HTML]{e1ecff}{12.2 $\pm$ 1.5} & \cellcolor[HTML]{e1ecff}{14.8 $\pm$ 2.6} & \cellcolor[HTML]{e1ecff}{5.3 $\pm$ 0.9} & \cellcolor[HTML]{e1ecff}{9.5 $\pm$ 1.6} & \cellcolor[HTML]{e1ecff}{3.8 $\pm$ 0.8} & \cellcolor[HTML]{e1ecff}{5.2 $\pm$ 1.4} \\
        \midrule
        SCoT~\cite{li2025structured} & GPT-5-mini & 13.8 $\pm$ 1.3 & 15.2 $\pm$ 1.3 & 4.7 $\pm$ 0.5 & 7.0 $\pm$ 0.6 & 4.8 $\pm$ 0.3 & 6.0 $\pm$ 0.5 & 14.6 $\pm$ 1.5 & 16.2 $\pm$ 1.2 & 5.9 $\pm$ 0.7 & 7.6 $\pm$ 1.2 & 5.2 $\pm$ 0.6 & 6.1 $\pm$ 1.0 & 11.3 $\pm$ 1.2 & 12.7 $\pm$ 0.9 & 4.5 $\pm$ 0.5 & 5.8 $\pm$ 0.9 & 3.9 $\pm$ 0.4 & 4.6 $\pm$ 0.7 \\
                              & Claude-Sonnet-3.7 & 15.2 $\pm$ 3.8 & 19.0 $\pm$ 2.7 & 9.2 $\pm$ 2.1 & 12.9 $\pm$ 1.5 & 6.0 $\pm$ 1.2 & 7.8 $\pm$ 0.6 & 17.6 $\pm$ 4.2 & 22.1 $\pm$ 3.1 & 10.9 $\pm$ 2.5 & 15.3 $\pm$ 1.8 & 7.1 $\pm$ 1.4 & 9.2 $\pm$ 0.7 & 12.4 $\pm$ 3.1 & 15.5 $\pm$ 2.2 & 7.3 $\pm$ 1.7 & 10.3 $\pm$ 1.2 & 4.8 $\pm$ 0.9 & 6.2 $\pm$ 0.5 \\
                              & Qwen3-Coder & 23.5 $\pm$ 6.1 & 24.4 $\pm$ 5.3 & 10.2 $\pm$ 1.5 & 15.5 $\pm$ 0.8 & 6.2 $\pm$ 0.7 & 8.9 $\pm$ 0.5 & 26.9 $\pm$ 6.8 & 27.9 $\pm$ 5.9 & 12.1 $\pm$ 1.8 & 18.3 $\pm$ 0.9 & 6.8 $\pm$ 0.9 & 10.5 $\pm$ 0.6 & 19.1 $\pm$ 5.0 & 19.9 $\pm$ 4.3 & 8.1 $\pm$ 1.2 & 12.4 $\pm$ 0.6 & 6.3 $\pm$ 0.6 & 7.1 $\pm$ 0.4 \\
        \midrule
        \cellcolor[HTML]{e1ecff}{OpenManus~\cite{openmanus2025}} & \cellcolor[HTML]{e1ecff}{GPT-5-mini} & \cellcolor[HTML]{e1ecff}{12.3 $\pm$ 0.9} & \cellcolor[HTML]{e1ecff}{15.4 $\pm$ 2.6} & \cellcolor[HTML]{e1ecff}{8.1 $\pm$ 2.0} & \cellcolor[HTML]{e1ecff}{12.2 $\pm$ 1.2} & \cellcolor[HTML]{e1ecff}{5.3 $\pm$ 1.2} & \cellcolor[HTML]{e1ecff}{5.6 $\pm$ 0.9} & \cellcolor[HTML]{e1ecff}{13.1 $\pm$ 1.0} & \cellcolor[HTML]{e1ecff}{16.3 $\pm$ 2.7} & \cellcolor[HTML]{e1ecff}{8.5 $\pm$ 2.1} & \cellcolor[HTML]{e1ecff}{12.8 $\pm$ 1.3} & \cellcolor[HTML]{e1ecff}{5.6 $\pm$ 1.3} & \cellcolor[HTML]{e1ecff}{5.9 $\pm$ 1.0} & \cellcolor[HTML]{e1ecff}{10.2 $\pm$ 0.7} & \cellcolor[HTML]{e1ecff}{12.8 $\pm$ 2.2} & \cellcolor[HTML]{e1ecff}{6.5 $\pm$ 1.6} & \cellcolor[HTML]{e1ecff}{9.8 $\pm$ 1.0} & \cellcolor[HTML]{e1ecff}{4.2 $\pm$ 1.0} & \cellcolor[HTML]{e1ecff}{4.5 $\pm$ 0.7} \\
        \cellcolor[HTML]{e1ecff}{} & \cellcolor[HTML]{e1ecff}{Claude-Sonnet-3.7} & \cellcolor[HTML]{e1ecff}{19.3 $\pm$ 0.3} & \cellcolor[HTML]{e1ecff}{28.2 $\pm$ 4.8} & \cellcolor[HTML]{e1ecff}{10.5 $\pm$ 1.7} & \cellcolor[HTML]{e1ecff}{15.4 $\pm$ 1.0} & \cellcolor[HTML]{e1ecff}{5.8 $\pm$ 0.4} & \cellcolor[HTML]{e1ecff}{8.1 $\pm$ 1.2} & \cellcolor[HTML]{e1ecff}{22.4 $\pm$ 0.4} & \cellcolor[HTML]{e1ecff}{32.8 $\pm$ 5.6} & \cellcolor[HTML]{e1ecff}{12.4 $\pm$ 2.0} & \cellcolor[HTML]{e1ecff}{18.2 $\pm$ 1.2} & \cellcolor[HTML]{e1ecff}{6.8 $\pm$ 0.5} & \cellcolor[HTML]{e1ecff}{9.6 $\pm$ 1.4} & \cellcolor[HTML]{e1ecff}{15.7 $\pm$ 0.2} & \cellcolor[HTML]{e1ecff}{23.0 $\pm$ 3.9} & \cellcolor[HTML]{e1ecff}{8.4 $\pm$ 1.4} & \cellcolor[HTML]{e1ecff}{12.3 $\pm$ 0.8} & \cellcolor[HTML]{e1ecff}{4.6 $\pm$ 0.3} & \cellcolor[HTML]{e1ecff}{6.5 $\pm$ 0.9} \\
        \cellcolor[HTML]{e1ecff}{} & \cellcolor[HTML]{e1ecff}{Qwen3-Coder} & \cellcolor[HTML]{e1ecff}{23.2 $\pm$ 3.9} & \cellcolor[HTML]{e1ecff}{25.5 $\pm$ 1.7} & \cellcolor[HTML]{e1ecff}{13.0 $\pm$ 1.8} & \cellcolor[HTML]{e1ecff}{17.8 $\pm$ 2.7} & \cellcolor[HTML]{e1ecff}{6.7 $\pm$ 0.9} & \cellcolor[HTML]{e1ecff}{9.3 $\pm$ 0.8} & \cellcolor[HTML]{e1ecff}{26.7 $\pm$ 4.5} & \cellcolor[HTML]{e1ecff}{29.3 $\pm$ 1.9} & \cellcolor[HTML]{e1ecff}{15.4 $\pm$ 2.1} & \cellcolor[HTML]{e1ecff}{21.1 $\pm$ 3.2} & \cellcolor[HTML]{e1ecff}{7.9 $\pm$ 1.0} & \cellcolor[HTML]{e1ecff}{11.0 $\pm$ 0.9} & \cellcolor[HTML]{e1ecff}{18.9 $\pm$ 3.2} & \cellcolor[HTML]{e1ecff}{20.8 $\pm$ 1.4} & \cellcolor[HTML]{e1ecff}{10.4 $\pm$ 1.4} & \cellcolor[HTML]{e1ecff}{14.2 $\pm$ 2.2} & \cellcolor[HTML]{e1ecff}{5.3 $\pm$ 0.7} & \cellcolor[HTML]{e1ecff}{7.4 $\pm$ 0.6} \\
        \midrule
        MetaGPT~\cite{hong2024metagpt} & GPT-5-mini & 12.6 $\pm$ 0.7 & 13.0 $\pm$ 1.8 & 6.6 $\pm$ 0.5 & 10.3 $\pm$ 1.1 & 4.0 $\pm$ 0.7 & 4.4 $\pm$ 0.6 & 13.3 $\pm$ 1.9 & 13.6 $\pm$ 2.5 & 7.5 $\pm$ 1.9 & 9.9 $\pm$ 1.6 & 4.5 $\pm$ 0.7 & 4.9 $\pm$ 0.5 & 10.3 $\pm$ 1.5 & 10.6 $\pm$ 2.0 & 5.7 $\pm$ 1.4 & 7.5 $\pm$ 1.2 & 3.4 $\pm$ 0.5 & 3.7 $\pm$ 0.3 \\
                                 & Claude-Sonnet-3.7 & 13.5 $\pm$ 3.4 & 16.9 $\pm$ 0.6 & 7.5 $\pm$ 1.9 & 12.4 $\pm$ 1.0 & 3.9 $\pm$ 0.2 & 4.6 $\pm$ 0.5 & 15.6 $\pm$ 3.9 & 19.5 $\pm$ 0.7 & 8.9 $\pm$ 2.2 & 14.7 $\pm$ 1.2 & 4.6 $\pm$ 0.2 & 5.4 $\pm$ 0.6 & 11.0 $\pm$ 2.8 & 13.8 $\pm$ 0.5 & 6.0 $\pm$ 1.5 & 9.9 $\pm$ 0.8 & 3.1 $\pm$ 0.2 & 3.7 $\pm$ 0.4 \\
                                 & Qwen3-Coder & 14.3 $\pm$ 2.0 & 22.2 $\pm$ 2.2 & 9.9 $\pm$ 0.8 & 11.6 $\pm$ 1.3 & 7.4 $\pm$ 0.6 & 8.8 $\pm$ 1.0 & 16.5 $\pm$ 2.3 & 25.6 $\pm$ 2.5 & 11.7 $\pm$ 0.9 & 13.7 $\pm$ 1.5 & 8.7 $\pm$ 0.7 & 10.4 $\pm$ 1.2 & 11.6 $\pm$ 1.6 & 18.1 $\pm$ 1.8 & 7.9 $\pm$ 0.6 & 9.3 $\pm$ 1.0 & 5.9 $\pm$ 0.5 & 7.0 $\pm$ 0.8 \\
        \midrule
        \cellcolor[HTML]{e1ecff}{DeepCode}~\cite{deepcode2025} & \cellcolor[HTML]{e1ecff}{GPT-5-mini} & \cellcolor[HTML]{e1ecff}{17.1 $\pm$ 3.8} & \cellcolor[HTML]{e1ecff}{17.9 $\pm$ 6.4} & \cellcolor[HTML]{e1ecff}{10.5 $\pm$ 1.3} & \cellcolor[HTML]{e1ecff}{14.2 $\pm$ 2.2} & \cellcolor[HTML]{e1ecff}{6.0 $\pm$ 1.0} & \cellcolor[HTML]{e1ecff}{6.4 $\pm$ 1.0} & \cellcolor[HTML]{e1ecff}{18.2 $\pm$ 4.0} & \cellcolor[HTML]{e1ecff}{19.0 $\pm$ 6.7} & \cellcolor[HTML]{e1ecff}{11.0 $\pm$ 1.4} & \cellcolor[HTML]{e1ecff}{14.9 $\pm$ 2.3} & \cellcolor[HTML]{e1ecff}{5.9 $\pm$ 1.6} & \cellcolor[HTML]{e1ecff}{6.7 $\pm$ 1.1} & \cellcolor[HTML]{e1ecff}{14.1 $\pm$ 3.1} & \cellcolor[HTML]{e1ecff}{14.8 $\pm$ 5.3} & \cellcolor[HTML]{e1ecff}{8.4 $\pm$ 1.0} & \cellcolor[HTML]{e1ecff}{11.4 $\pm$ 1.8} & \cellcolor[HTML]{e1ecff}{5.3 $\pm$ 0.8} & \cellcolor[HTML]{e1ecff}{6.0 $\pm$ 0.4} \\
        \cellcolor[HTML]{e1ecff}{}& \cellcolor[HTML]{e1ecff}{Claude-Sonnet-3.7} & \cellcolor[HTML]{e1ecff}{29.4 $\pm$ 3.8} & \cellcolor[HTML]{e1ecff}{32.3 $\pm$ 4.6} & \cellcolor[HTML]{e1ecff}{15.6 $\pm$ 1.4} & \cellcolor[HTML]{e1ecff}{17.2 $\pm$ 2.6} & \cellcolor[HTML]{e1ecff}{5.3 $\pm$ 0.5} & \cellcolor[HTML]{e1ecff}{10.1 $\pm$ 1.1} & \cellcolor[HTML]{e1ecff}{34.1 $\pm$ 4.4} & \cellcolor[HTML]{e1ecff}{37.5 $\pm$ 5.3} & \cellcolor[HTML]{e1ecff}{18.4 $\pm$ 1.6} & \cellcolor[HTML]{e1ecff}{20.3 $\pm$ 3.0} & \cellcolor[HTML]{e1ecff}{6.2 $\pm$ 0.6} & \cellcolor[HTML]{e1ecff}{11.9 $\pm$ 1.3} & \cellcolor[HTML]{e1ecff}{24.0 $\pm$ 3.1} & \cellcolor[HTML]{e1ecff}{26.4 $\pm$ 3.8} & \cellcolor[HTML]{e1ecff}{12.5 $\pm$ 1.1} & \cellcolor[HTML]{e1ecff}{13.8 $\pm$ 2.1} & \cellcolor[HTML]{e1ecff}{4.2 $\pm$ 0.4} & \cellcolor[HTML]{e1ecff}{8.1 $\pm$ 0.9} \\
        \cellcolor[HTML]{e1ecff}{}& \cellcolor[HTML]{e1ecff}{Qwen3-Coder} & \cellcolor[HTML]{e1ecff}{27.4 $\pm$ 6.3} & \cellcolor[HTML]{e1ecff}{37.6 $\pm$ 6.3} & \cellcolor[HTML]{e1ecff}{13.3 $\pm$ 1.2} & \cellcolor[HTML]{e1ecff}{19.4 $\pm$ 2.9} & \cellcolor[HTML]{e1ecff}{4.9 $\pm$ 1.1} & \cellcolor[HTML]{e1ecff}{9.8 $\pm$ 1.4} & \cellcolor[HTML]{e1ecff}{31.6 $\pm$ 7.3} & \cellcolor[HTML]{e1ecff}{43.4 $\pm$ 7.3} & \cellcolor[HTML]{e1ecff}{15.8 $\pm$ 1.4} & \cellcolor[HTML]{e1ecff}{23.0 $\pm$ 3.4} & \cellcolor[HTML]{e1ecff}{5.8 $\pm$ 1.3} & \cellcolor[HTML]{e1ecff}{11.6 $\pm$ 1.6} & \cellcolor[HTML]{e1ecff}{22.4 $\pm$ 5.1} & \cellcolor[HTML]{e1ecff}{30.7 $\pm$ 5.1} & \cellcolor[HTML]{e1ecff}{10.6 $\pm$ 0.9} & \cellcolor[HTML]{e1ecff}{15.5 $\pm$ 2.3} & \cellcolor[HTML]{e1ecff}{3.9 $\pm$ 0.9} & \cellcolor[HTML]{e1ecff}{7.8 $\pm$ 1.1} \\
        \midrule
        \cellcolor[HTML]{e1dcef}{\textit{\textbf{\mytitle}}} & \cellcolor[HTML]{e1dcef}{GPT-5-mini} & \cellcolor[HTML]{e1dcef}{23.0 $\pm$ 1.4} & \cellcolor[HTML]{e1dcef}{26.8 $\pm$ 2.4} & \cellcolor[HTML]{e1dcef}{12.7 $\pm$ 1.1} & \cellcolor[HTML]{e1dcef}{19.7 $\pm$ 2.1} & \cellcolor[HTML]{e1dcef}{8.3 $\pm$ 1.3} & \cellcolor[HTML]{e1dcef}{9.8 $\pm$ 2.1} & \cellcolor[HTML]{e1dcef}{26.5 $\pm$ 1.6} & \cellcolor[HTML]{e1dcef}{30.9 $\pm$ 2.8} & \cellcolor[HTML]{e1dcef}{15.0 $\pm$ 1.3} & \cellcolor[HTML]{e1dcef}{23.3 $\pm$ 2.5} & \cellcolor[HTML]{e1dcef}{9.8 $\pm$ 1.5} & \cellcolor[HTML]{e1dcef}{11.6 $\pm$ 2.5} & \cellcolor[HTML]{e1dcef}{18.7 $\pm$ 1.1} & \cellcolor[HTML]{e1dcef}{21.8 $\pm$ 2.0} & \cellcolor[HTML]{e1dcef}{10.2 $\pm$ 0.9} & \cellcolor[HTML]{e1dcef}{15.8 $\pm$ 1.7} & \cellcolor[HTML]{e1dcef}{6.6 $\pm$ 1.0} & \cellcolor[HTML]{e1dcef}{7.8 $\pm$ 1.7} \\
        \cellcolor[HTML]{e1dcef}{} & \cellcolor[HTML]{e1dcef}{Claude-Sonnet-3.7} & \cellcolor[HTML]{e1dcef}{29.5 $\pm$ 2.7} & \cellcolor[HTML]{e1dcef}{35.6 $\pm$ 4.4} & \cellcolor[HTML]{e1dcef}{15.5 $\pm$ 0.6} & \cellcolor[HTML]{e1dcef}{19.4 $\pm$ 2.9} & \cellcolor[HTML]{e1dcef}{13.9 $\pm$ 0.9} & \cellcolor[HTML]{e1dcef}{17.4 $\pm$ 1.3} & \cellcolor[HTML]{e1dcef}{34.1 $\pm$ 3.1} & \cellcolor[HTML]{e1dcef}{41.2 $\pm$ 5.1} & \cellcolor[HTML]{e1dcef}{18.3 $\pm$ 0.7} & \cellcolor[HTML]{e1dcef}{23.0 $\pm$ 3.4} & \cellcolor[HTML]{e1dcef}{16.4 $\pm$ 1.1} & \cellcolor[HTML]{e1dcef}{20.6 $\pm$ 1.5} & \cellcolor[HTML]{e1dcef}{24.0 $\pm$ 2.2} & \cellcolor[HTML]{e1dcef}{29.0 $\pm$ 3.6} & \cellcolor[HTML]{e1dcef}{12.4 $\pm$ 0.5} & \cellcolor[HTML]{e1dcef}{15.5 $\pm$ 2.3} & \cellcolor[HTML]{e1dcef}{11.1 $\pm$ 0.7} & \cellcolor[HTML]{e1dcef}{13.9 $\pm$ 1.0} \\
        \cellcolor[HTML]{e1dcef}{} & \cellcolor[HTML]{e1dcef}{Qwen3-Coder} & \cellcolor[HTML]{e1dcef}{\textbf{32.4 $\pm$ 5.8}} & \cellcolor[HTML]{e1dcef}{\textbf{38.1 $\pm$ 4.9}} & \cellcolor[HTML]{e1dcef}{\textbf{17.6 $\pm$ 2.1}} & \cellcolor[HTML]{e1dcef}{\textbf{22.0 $\pm$ 3.2}} & \cellcolor[HTML]{e1dcef}{\textbf{16.0 $\pm$ 1.1}} & \cellcolor[HTML]{e1dcef}{\textbf{18.9 $\pm$ 4.0}} & \cellcolor[HTML]{e1dcef}{\textbf{37.4 $\pm$ 6.7}} & \cellcolor[HTML]{e1dcef}{\textbf{44.0 $\pm$ 5.7}} & \cellcolor[HTML]{e1dcef}{\textbf{20.8 $\pm$ 2.5}} & \cellcolor[HTML]{e1dcef}{\textbf{26.0 $\pm$ 3.8}} & \cellcolor[HTML]{e1dcef}{\textbf{18.9 $\pm$ 1.3}} & \cellcolor[HTML]{e1dcef}{\textbf{22.4 $\pm$ 4.7}} & \cellcolor[HTML]{e1dcef}{\textbf{26.4 $\pm$ 4.7}} & \cellcolor[HTML]{e1dcef}{\textbf{31.0 $\pm$ 4.0}} & \cellcolor[HTML]{e1dcef}{\textbf{14.1 $\pm$ 1.7}} & \cellcolor[HTML]{e1dcef}{\textbf{17.6 $\pm$ 2.6}} & \cellcolor[HTML]{e1dcef}{\textbf{12.8 $\pm$ 0.9}} & \cellcolor[HTML]{e1dcef}{\textbf{15.1 $\pm$ 3.2}} \\
        \bottomrule
        \end{tabular}
    }
    \label{tab:rq2}
\end{table*}

\intuition{
{\bf Answer to RQ-1}: \mytitle outperforms all baseline methods across LLMs and languages.
\mytitle also shows clear advantages over other multi-agent baselines.
These results demonstrate that existing prompt-engineering strategies and multi-agent approaches 
remain limited for GUI application generation.
The performance differences highlight the necessity of combining dynamic GUI testing with iterative repair capabilities from visual signals and dynamic interaction.
}
\label{sec:rq1}

\subsection{RQ-2: How does the efficiency of \mytitle compare with the baselines?}

\noindent
\textbf{Objective.}
This research question evaluates the computational efficiency and resource consumption of \mytitle compared to baseline methods.
Beyond effectiveness, computational efficiency considers practical deployment of GUI code generation systems.
We investigate token consumption, processing time, and cost-effectiveness ratios across different methods and models to assess the trade-offs between enhancement sophistication and computational overhead.

\noindent
\textbf{Experimental Design.}
We conduct efficiency analysis across representative methods and pure LLM to ensure statistical reliability.
For each method, we measure: (1) \textit{Total token consumption} from API calls during code generation, 
(2) \textit{Processing time} from task initiation to completion, 
and (3) \textit{Per-function metrics} to normalize for workload differences.

\begin{table*}[htbp!]
    \centering
    \caption{\revise{Computational efficiency comparison across baselines and \mytitle (GPT-5-mini as backbone LLM).
    Efficiency@k calculated as Play@k divided by tokens per function ($\times 10^{3}$). 
    We report mean values over 5 independent runs with 95\% confidence intervals calculated using the Student's $t$-distribution.}}
    \resizebox{\linewidth}{!}
    {
        \begin{tabular}{l|c|c|c|c|c|c}
        \toprule
        \textbf{LLMs / Methods} & \textbf{Tokens $\downarrow$} & 
        \textbf{Tokens / Func $\downarrow$} & \textbf{Play@1$\uparrow$} & \textbf{Play@3$\uparrow$} & \textbf{Efficiency@1$\uparrow$} & \textbf{Efficiency@3$\uparrow$} \\
        \midrule
        GPT-5-mini & \textbf{128K $\pm$ 7K} & \textbf{4267 $\pm$ 239} & 4.3 $\pm$ 0.5 & 5.2 $\pm$ 0.6 & 1.01 $\pm$ 0.13 & 1.22 $\pm$ 0.16 \\
        \midrule
        \cellcolor[HTML]{e1ecff}{SCoT} & \cellcolor[HTML]{e1ecff}{184K $\pm$ 23K} & \cellcolor[HTML]{e1ecff}{6135 $\pm$ 764} & \cellcolor[HTML]{e1ecff}{4.8 $\pm$ 0.3} & \cellcolor[HTML]{e1ecff}{6.0 $\pm$ 0.5} & \cellcolor[HTML]{e1ecff}{0.78 $\pm$ 0.11} & \cellcolor[HTML]{e1ecff}{0.98 $\pm$ 0.15} \\
        HCPCoder & 373K $\pm$ 63K & 12422 $\pm$ 2105 & 0.3 $\pm$ 0.0 & 0.3 $\pm$ 0.1 & 0.02 $\pm$ 0.00 & 0.02 $\pm$ 0.01 \\
        \cellcolor[HTML]{e1ecff}{MetaGPT} & \cellcolor[HTML]{e1ecff}{148K $\pm$ 12K} & \cellcolor[HTML]{e1ecff}{4931 $\pm$ 416} & \cellcolor[HTML]{e1ecff}{4.0 $\pm$ 0.7} & \cellcolor[HTML]{e1ecff}{4.4 $\pm$ 0.6} & \cellcolor[HTML]{e1ecff}{0.81 $\pm$ 0.16} & \cellcolor[HTML]{e1ecff}{0.89 $\pm$ 0.14} \\
        OpenManus & 176K $\pm$ 23K & 5869 $\pm$ 771 & 5.3 $\pm$ 0.8 & 5.8 $\pm$ 0.3 & 0.90 $\pm$ 0.19 & 0.99 $\pm$ 0.14 \\
        \cellcolor[HTML]{e1ecff}{DeepCode} & \cellcolor[HTML]{e1ecff}{252K $\pm$ 28K} & \cellcolor[HTML]{e1ecff}{8406 $\pm$ 939} & \cellcolor[HTML]{e1ecff}{6.0 $\pm$ 1.0} & \cellcolor[HTML]{e1ecff}{6.4 $\pm$ 1.0} & \cellcolor[HTML]{e1ecff}{0.72 $\pm$ 0.15} & \cellcolor[HTML]{e1ecff}{0.76 $\pm$ 0.15} \\
        \cellcolor[HTML]{e1dcef}{\textbf{\textit{\mytitle}}} & \cellcolor[HTML]{e1dcef}{164K $\pm$ 21K} & \cellcolor[HTML]{e1dcef}{5480 $\pm$ 686} & \cellcolor[HTML]{e1dcef}{\textbf{8.3 $\pm$ 1.3}} & \cellcolor[HTML]{e1dcef}{\textbf{9.8 $\pm$ 2.1}} & \cellcolor[HTML]{e1dcef}{\textbf{1.51 $\pm$ 0.30}} & \cellcolor[HTML]{e1dcef}{\textbf{1.79 $\pm$ 0.44}} \\
        \bottomrule
        \end{tabular}
    }
    \label{tab:rq3_new}
\end{table*}

\noindent
\textbf{Results.}
Table~\ref{tab:rq3_new} presents efficiency metrics across methods and models, 
revealing significant trade-offs between enhancement sophistication and computational overhead.
GPT-5-mini consumes 128K tokens with 4,267 tokens per function, achieving 4.3\% Play@1 and 1.01 Efficiency@1.
The base model shows moderate token consumption with reasonable behavioral validation performance.
Enhancement methods exhibit varying efficiency characteristics compared to GPT-5-mini.
SCoT increases token usage by 44\% (184K vs 128K tokens) but achieves marginal improvement in Play@1 (4.8\% vs 4.3\%), resulting in reduced efficiency (0.98 vs 1.22).
HCPCoder presents the most concerning efficiency profile, consuming 191\% more tokens (373K vs 128K) while achieving catastrophically poor behavioral validation (0.3\% Play@1), resulting in extremely low Efficiency@1 (0.02), attributed to its poor repository context management.
MetaGPT demonstrates moderate token efficiency with 148K tokens, maintaining 4.0\% Play@1 performance with 0.89 efficiency.
OpenManus requires substantial computational resources (176K tokens) but achieves competitive Play@1 performance (5.3\%), resulting in moderate Efficiency@1 of 0.90.
DeepCode shows the best behavioral validation among baselines (6.0\% Play@1) but high token consumption (252K tokens) limits its Efficiency@1 to 0.72.
\mytitle demonstrates the 3rd best cost-effectiveness characteristics among all evaluated agentic methods.
The framework consumes 164K tokens for 30 functions, achieving 5,480 tokens per function with 8.3\% Play@1 performance. 
\revise{Our preliminary analysis on 164K tokens (Table~\ref{tab:rq3_new}) indicates that the Testing Phase accounts for \textasciitilde22\% consumption, 
while the Generation Phase (coding, iterative patching) accounts for \textasciitilde78\%.}
\mytitle delivers significantly better performance per token consumed compared to all baseline methods.

\intuition{
{\bf Answer to RQ-2}: \mytitle demonstrates the 3rd best cost-effectiveness (Token consumption) among all evaluated methods, 
achieving the highest Efficiency@k, and Play@k performance.
These results demonstrate that \mytitle provides acceptable resource utilization 
and behavioral validation performance for practical deployment scenarios.
}
\label{sec:rq2}

\subsection{RQ-3: How do different components and models affect \mytitle's effectiveness?}

\noindent
\textbf{Objective.}
This research question examines the contribution of each component within the \mytitle framework and evaluates its model-agnostic effectiveness across diverse LLM architectures.
We investigate how GUI feedback, APR, and their integration impact performance across Exec@k, Pass@k, and Play@k.
Additionally, we assess whether \mytitle provides consistent benefits across different LLMs with varying capabilities, architectural designs, and parameter scales.

\begin{table*}[htbp!]
    \centering
    \caption{The performance (\%) of \mytitle under different configurations, 5 runs + 95\% CI (t-distribution).}
    \resizebox{1.0\linewidth}{!}
    {
        \begin{tabular}{l|ccc|cccccc|cccccc|cccccc}
        \toprule
        \multirow{2}{*}{\textbf{LLMs}} & \multirow{2}{*}{\textbf{APR}} & \multirow{2}{*}{\textbf{\makecell{GUI}}} & \multirow{2}{*}{\textbf{\makecell{Context}}} & \multicolumn{6}{c|}{\textbf{Python}} & \multicolumn{6}{c|}{\textbf{JavaScript}} & \multicolumn{6}{c}{\textbf{TypeScript}} \\
        \cmidrule(lr){5-10} \cmidrule(lr){11-16} \cmidrule(lr){17-22}
        & & & & \textbf{Exec@1} & \textbf{Exec@3} & \textbf{Pass@1} & \textbf{Pass@3} & \textbf{Play@1} & \textbf{Play@3} & \textbf{Exec@1} & \textbf{Exec@3} & \textbf{Pass@1} & \textbf{Pass@3} & \textbf{Play@1} & \textbf{Play@3} & \textbf{Exec@1} & \textbf{Exec@3} & \textbf{Pass@1} & \textbf{Pass@3} & \textbf{Play@1} & \textbf{Play@3} \\
        \midrule
        \cellcolor[HTML]{e1ecff}{GPT-5-mini} & \cellcolor[HTML]{e1ecff}{\ding{51}} & \cellcolor[HTML]{e1ecff}{\ding{51}} & \cellcolor[HTML]{e1ecff}{\ding{51}} & \cellcolor[HTML]{e1ecff}{23.0 $\pm$ 1.4} & \cellcolor[HTML]{e1ecff}{26.8 $\pm$ 2.4} & \cellcolor[HTML]{e1ecff}{12.7 $\pm$ 1.1} & \cellcolor[HTML]{e1ecff}{19.7 $\pm$ 2.1} & \cellcolor[HTML]{e1ecff}{8.3 $\pm$ 1.3} & \cellcolor[HTML]{e1ecff}{9.8 $\pm$ 2.1} & \cellcolor[HTML]{e1ecff}{26.5 $\pm$ 1.6} & \cellcolor[HTML]{e1ecff}{30.9 $\pm$ 2.8} & \cellcolor[HTML]{e1ecff}{15.0 $\pm$ 1.3} & \cellcolor[HTML]{e1ecff}{23.3 $\pm$ 2.5} & \cellcolor[HTML]{e1ecff}{9.8 $\pm$ 1.5} & \cellcolor[HTML]{e1ecff}{11.6 $\pm$ 2.5} & \cellcolor[HTML]{e1ecff}{18.7 $\pm$ 1.1} & \cellcolor[HTML]{e1ecff}{21.8 $\pm$ 2.0} & \cellcolor[HTML]{e1ecff}{10.2 $\pm$ 0.9} & \cellcolor[HTML]{e1ecff}{15.8 $\pm$ 1.7} & \cellcolor[HTML]{e1ecff}{6.6 $\pm$ 1.0} & \cellcolor[HTML]{e1ecff}{7.8 $\pm$ 1.7} \\
        \cellcolor[HTML]{e1ecff}{} & \cellcolor[HTML]{e1ecff}{\ding{55}} & \cellcolor[HTML]{e1ecff}{\ding{51}} & \cellcolor[HTML]{e1ecff}{\ding{51}} & \cellcolor[HTML]{e1ecff}{15.5 $\pm$ 1.1} & \cellcolor[HTML]{e1ecff}{18.3 $\pm$ 2.3} & \cellcolor[HTML]{e1ecff}{6.8 $\pm$ 0.4} & \cellcolor[HTML]{e1ecff}{9.2 $\pm$ 0.5} & \cellcolor[HTML]{e1ecff}{4.9 $\pm$ 0.3} & \cellcolor[HTML]{e1ecff}{7.9 $\pm$ 0.8} & \cellcolor[HTML]{e1ecff}{17.9 $\pm$ 1.3} & \cellcolor[HTML]{e1ecff}{21.1 $\pm$ 2.7} & \cellcolor[HTML]{e1ecff}{7.7 $\pm$ 0.5} & \cellcolor[HTML]{e1ecff}{13.5 $\pm$ 0.6} & \cellcolor[HTML]{e1ecff}{5.8 $\pm$ 0.4} & \cellcolor[HTML]{e1ecff}{9.3 $\pm$ 0.9} & \cellcolor[HTML]{e1ecff}{12.6 $\pm$ 0.9} & \cellcolor[HTML]{e1ecff}{14.9 $\pm$ 1.9} & \cellcolor[HTML]{e1ecff}{4.8 $\pm$ 0.3} & \cellcolor[HTML]{e1ecff}{7.8 $\pm$ 0.4} & \cellcolor[HTML]{e1ecff}{3.9 $\pm$ 0.2} & \cellcolor[HTML]{e1ecff}{6.3 $\pm$ 0.6} \\
        \cellcolor[HTML]{e1ecff}{} & \cellcolor[HTML]{e1ecff}{\ding{51}} & \cellcolor[HTML]{e1ecff}{\ding{55}} & \cellcolor[HTML]{e1ecff}{\ding{51}} & \cellcolor[HTML]{e1ecff}{20.5 $\pm$ 4.1} & \cellcolor[HTML]{e1ecff}{24.1 $\pm$ 5.9} & \cellcolor[HTML]{e1ecff}{10.4 $\pm$ 1.2} & \cellcolor[HTML]{e1ecff}{15.4 $\pm$ 1.0} & \cellcolor[HTML]{e1ecff}{3.0 $\pm$ 0.4} & \cellcolor[HTML]{e1ecff}{5.3 $\pm$ 1.0} & \cellcolor[HTML]{e1ecff}{23.6 $\pm$ 4.7} & \cellcolor[HTML]{e1ecff}{27.9 $\pm$ 6.8} & \cellcolor[HTML]{e1ecff}{12.3 $\pm$ 1.4} & \cellcolor[HTML]{e1ecff}{18.2 $\pm$ 1.2} & \cellcolor[HTML]{e1ecff}{3.6 $\pm$ 0.5} & \cellcolor[HTML]{e1ecff}{6.3 $\pm$ 1.2} & \cellcolor[HTML]{e1ecff}{16.7 $\pm$ 3.3} & \cellcolor[HTML]{e1ecff}{19.6 $\pm$ 4.8} & \cellcolor[HTML]{e1ecff}{8.3 $\pm$ 1.0} & \cellcolor[HTML]{e1ecff}{12.3 $\pm$ 0.8} & \cellcolor[HTML]{e1ecff}{2.4 $\pm$ 0.3} & \cellcolor[HTML]{e1ecff}{4.2 $\pm$ 0.8} \\
        \cellcolor[HTML]{e1ecff}{} & \cellcolor[HTML]{e1ecff}{\ding{55}} & \cellcolor[HTML]{e1ecff}{\ding{55}} & \cellcolor[HTML]{e1ecff}{\ding{51}} & \cellcolor[HTML]{e1ecff}{12.8 $\pm$ 2.8} & \cellcolor[HTML]{e1ecff}{14.3 $\pm$ 2.6} & \cellcolor[HTML]{e1ecff}{5.8 $\pm$ 1.0} & \cellcolor[HTML]{e1ecff}{7.0 $\pm$ 1.0} & \cellcolor[HTML]{e1ecff}{4.5 $\pm$ 0.5} & \cellcolor[HTML]{e1ecff}{4.9 $\pm$ 0.3} & \cellcolor[HTML]{e1ecff}{15.1 $\pm$ 3.2} & \cellcolor[HTML]{e1ecff}{16.5 $\pm$ 3.0} & \cellcolor[HTML]{e1ecff}{6.8 $\pm$ 1.2} & \cellcolor[HTML]{e1ecff}{8.3 $\pm$ 1.2} & \cellcolor[HTML]{e1ecff}{5.3 $\pm$ 0.6} & \cellcolor[HTML]{e1ecff}{5.8 $\pm$ 0.4} & \cellcolor[HTML]{e1ecff}{11.0 $\pm$ 2.3} & \cellcolor[HTML]{e1ecff}{11.6 $\pm$ 2.1} & \cellcolor[HTML]{e1ecff}{4.6 $\pm$ 0.8} & \cellcolor[HTML]{e1ecff}{5.6 $\pm$ 0.8} & \cellcolor[HTML]{e1ecff}{3.6 $\pm$ 0.4} & \cellcolor[HTML]{e1ecff}{3.9 $\pm$ 0.2} \\
        \cellcolor[HTML]{e1ecff}{} & \cellcolor[HTML]{e1ecff}{\ding{55}} & \cellcolor[HTML]{e1ecff}{\ding{55}} & \cellcolor[HTML]{e1ecff}{\ding{55}} & \cellcolor[HTML]{e1ecff}{12.6 $\pm$ 1.6} & \cellcolor[HTML]{e1ecff}{14.3 $\pm$ 1.0} & \cellcolor[HTML]{e1ecff}{5.4 $\pm$ 0.9} & \cellcolor[HTML]{e1ecff}{7.1 $\pm$ 0.7} & \cellcolor[HTML]{e1ecff}{4.2 $\pm$ 0.4} & \cellcolor[HTML]{e1ecff}{4.9 $\pm$ 0.4} & \cellcolor[HTML]{e1ecff}{14.5 $\pm$ 1.8} & \cellcolor[HTML]{e1ecff}{16.5 $\pm$ 1.2} & \cellcolor[HTML]{e1ecff}{6.4 $\pm$ 1.1} & \cellcolor[HTML]{e1ecff}{8.4 $\pm$ 0.8} & \cellcolor[HTML]{e1ecff}{5.0 $\pm$ 0.5} & \cellcolor[HTML]{e1ecff}{5.3 $\pm$ 0.6} & \cellcolor[HTML]{e1ecff}{10.2 $\pm$ 1.3} & \cellcolor[HTML]{e1ecff}{11.6 $\pm$ 0.8} & \cellcolor[HTML]{e1ecff}{4.3 $\pm$ 0.7} & \cellcolor[HTML]{e1ecff}{5.7 $\pm$ 0.6} & \cellcolor[HTML]{e1ecff}{3.4 $\pm$ 0.3} & \cellcolor[HTML]{e1ecff}{3.9 $\pm$ 0.3} \\
        \midrule
        Claude-3.7 & \ding{51} & \ding{51} & \ding{51} & {29.5 $\pm$ 2.7} & {35.6 $\pm$ 4.4} & {15.5 $\pm$ 0.6} & {19.4 $\pm$ 2.9} & {13.9 $\pm$ 0.9} & {17.4 $\pm$ 1.3} & {34.1 $\pm$ 3.1} & {41.2 $\pm$ 5.1} & {18.3 $\pm$ 0.7} & {23.0 $\pm$ 3.4} & {16.4 $\pm$ 1.1} & {20.6 $\pm$ 1.5} & {24.0 $\pm$ 2.2} & {29.0 $\pm$ 3.6} & {12.4 $\pm$ 0.5} & {15.5 $\pm$ 2.3} & {11.1 $\pm$ 0.7} & {13.9 $\pm$ 1.0} \\
        & \ding{55} & \ding{51} & \ding{51} & 19.8 $\pm$ 2.3 & 21.5 $\pm$ 3.1 & 8.2 $\pm$ 0.7 & 12.4 $\pm$ 1.0 & 8.7 $\pm$ 2.0 & 9.7 $\pm$ 0.7 & 22.9 $\pm$ 2.7 & 24.8 $\pm$ 3.6 & 9.7 $\pm$ 0.8 & 14.7 $\pm$ 1.2 & 8.3 $\pm$ 2.4 & 11.5 $\pm$ 0.8 & 16.1 $\pm$ 1.9 & 17.5 $\pm$ 2.5 & 8.6 $\pm$ 0.6 & 9.9 $\pm$ 0.8 & 6.9 $\pm$ 1.6 & 7.8 $\pm$ 0.6 \\
        & \ding{51} & \ding{55} & \ding{51} & 29.1 $\pm$ 5.6 & 34.1 $\pm$ 4.1 & 13.0 $\pm$ 2.3 & 16.5 $\pm$ 4.1 & 5.3 $\pm$ 0.7 & 6.9 $\pm$ 0.9 & 33.7 $\pm$ 6.5 & 39.4 $\pm$ 4.7 & 15.4 $\pm$ 2.7 & 19.5 $\pm$ 4.8 & 6.3 $\pm$ 0.8 & 8.2 $\pm$ 1.1 & 23.7 $\pm$ 4.6 & 27.7 $\pm$ 3.3 & 10.4 $\pm$ 1.8 & 13.2 $\pm$ 3.3 & 4.2 $\pm$ 0.6 & 5.5 $\pm$ 0.7 \\
        & \ding{55} & \ding{55} & \ding{51} & 16.4 $\pm$ 2.5 & 17.2 $\pm$ 2.2 & 5.3 $\pm$ 0.5 & 7.0 $\pm$ 0.7 & 5.0 $\pm$ 0.2 & 5.9 $\pm$ 0.9 & 19.0 $\pm$ 2.9 & 19.9 $\pm$ 2.5 & 6.3 $\pm$ 0.6 & 8.3 $\pm$ 0.8 & 5.9 $\pm$ 0.2 & 7.0 $\pm$ 1.1 & 13.3 $\pm$ 2.0 & 14.0 $\pm$ 1.8 & 4.2 $\pm$ 0.4 & 5.6 $\pm$ 0.6 & 4.0 $\pm$ 0.2 & 4.7 $\pm$ 0.7 \\
        & \ding{55} & \ding{55} & \ding{55} & 12.4 $\pm$ 1.0 & 13.7 $\pm$ 2.1 & 5.8 $\pm$ 0.7 & 9.0 $\pm$ 1.2 & 4.1 $\pm$ 0.8 & 7.6 $\pm$ 0.9 & 14.3 $\pm$ 1.2 & 15.8 $\pm$ 2.4 & 6.9 $\pm$ 0.8 & 10.7 $\pm$ 1.4 & 4.9 $\pm$ 0.9 & 9.0 $\pm$ 1.1 & 10.1 $\pm$ 0.8 & 11.1 $\pm$ 1.7 & 4.6 $\pm$ 0.6 & 7.2 $\pm$ 1.0 & 3.3 $\pm$ 0.6 & 6.1 $\pm$ 0.7 \\
        \midrule
        \cellcolor[HTML]{e1ecff}{Qwen3-Coder} & \cellcolor[HTML]{e1ecff}{\ding{51}} & \cellcolor[HTML]{e1ecff}{\ding{51}} & \cellcolor[HTML]{e1ecff}{\ding{51}} & \cellcolor[HTML]{e1ecff}{\textbf{32.4 $\pm$ 5.8}} & \cellcolor[HTML]{e1ecff}{\textbf{38.1 $\pm$ 4.9}} & \cellcolor[HTML]{e1ecff}{\textbf{17.6 $\pm$ 2.1}} & \cellcolor[HTML]{e1ecff}{\textbf{22.0 $\pm$ 3.2}} & \cellcolor[HTML]{e1ecff}{\textbf{16.0 $\pm$ 1.1}} & \cellcolor[HTML]{e1ecff}{\textbf{18.9 $\pm$ 4.0}} & \cellcolor[HTML]{e1ecff}{\textbf{37.4 $\pm$ 6.7}} & \cellcolor[HTML]{e1ecff}{\textbf{44.0 $\pm$ 5.7}} & \cellcolor[HTML]{e1ecff}{\textbf{20.8 $\pm$ 2.5}} & \cellcolor[HTML]{e1ecff}{\textbf{26.0 $\pm$ 3.8}} & \cellcolor[HTML]{e1ecff}{\textbf{18.9 $\pm$ 1.3}} & \cellcolor[HTML]{e1ecff}{\textbf{22.4 $\pm$ 4.7}} & \cellcolor[HTML]{e1ecff}{\textbf{26.4 $\pm$ 4.7}} & \cellcolor[HTML]{e1ecff}{\textbf{31.0 $\pm$ 4.0}} & \cellcolor[HTML]{e1ecff}{\textbf{14.1 $\pm$ 1.7}} & \cellcolor[HTML]{e1ecff}{\textbf{17.6 $\pm$ 2.6}} & \cellcolor[HTML]{e1ecff}{\textbf{12.8 $\pm$ 0.9}} & \cellcolor[HTML]{e1ecff}{\textbf{15.1 $\pm$ 3.2}} \\
        \cellcolor[HTML]{e1ecff}{} & \cellcolor[HTML]{e1ecff}{\ding{55}} & \cellcolor[HTML]{e1ecff}{\ding{51}} & \cellcolor[HTML]{e1ecff}{\ding{51}} & \cellcolor[HTML]{e1ecff}{20.8 $\pm$ 3.0} & \cellcolor[HTML]{e1ecff}{24.1 $\pm$ 3.5} & \cellcolor[HTML]{e1ecff}{10.4 $\pm$ 0.9} & \cellcolor[HTML]{e1ecff}{15.1 $\pm$ 1.6} & \cellcolor[HTML]{e1ecff}{7.7 $\pm$ 1.6} & \cellcolor[HTML]{e1ecff}{10.9 $\pm$ 2.0} & \cellcolor[HTML]{e1ecff}{24.0 $\pm$ 3.5} & \cellcolor[HTML]{e1ecff}{27.9 $\pm$ 4.0} & \cellcolor[HTML]{e1ecff}{12.3 $\pm$ 1.1} & \cellcolor[HTML]{e1ecff}{17.8 $\pm$ 1.9} & \cellcolor[HTML]{e1ecff}{9.1 $\pm$ 1.9} & \cellcolor[HTML]{e1ecff}{14.6 $\pm$ 2.4} & \cellcolor[HTML]{e1ecff}{16.9 $\pm$ 2.4} & \cellcolor[HTML]{e1ecff}{19.6 $\pm$ 2.8} & \cellcolor[HTML]{e1ecff}{8.3 $\pm$ 0.7} & \cellcolor[HTML]{e1ecff}{12.1 $\pm$ 1.3} & \cellcolor[HTML]{e1ecff}{6.2 $\pm$ 1.3} & \cellcolor[HTML]{e1ecff}{9.8 $\pm$ 1.6} \\
        \cellcolor[HTML]{e1ecff}{} & \cellcolor[HTML]{e1ecff}{\ding{51}} & \cellcolor[HTML]{e1ecff}{\ding{55}} & \cellcolor[HTML]{e1ecff}{\ding{51}} & \cellcolor[HTML]{e1ecff}{31.3 $\pm$ 5.3} & \cellcolor[HTML]{e1ecff}{34.2 $\pm$ 3.0} & \cellcolor[HTML]{e1ecff}{9.7 $\pm$ 2.5} & \cellcolor[HTML]{e1ecff}{14.8 $\pm$ 0.8} & \cellcolor[HTML]{e1ecff}{7.5 $\pm$ 1.0} & \cellcolor[HTML]{e1ecff}{9.3 $\pm$ 3.0} & \cellcolor[HTML]{e1ecff}{36.2 $\pm$ 6.1} & \cellcolor[HTML]{e1ecff}{39.5 $\pm$ 3.5} & \cellcolor[HTML]{e1ecff}{11.5 $\pm$ 2.9} & \cellcolor[HTML]{e1ecff}{17.5 $\pm$ 0.9} & \cellcolor[HTML]{e1ecff}{8.9 $\pm$ 1.2} & \cellcolor[HTML]{e1ecff}{10.5 $\pm$ 3.5} & \cellcolor[HTML]{e1ecff}{25.5 $\pm$ 4.3} & \cellcolor[HTML]{e1ecff}{27.8 $\pm$ 2.4} & \cellcolor[HTML]{e1ecff}{7.8 $\pm$ 2.0} & \cellcolor[HTML]{e1ecff}{11.8 $\pm$ 0.6} & \cellcolor[HTML]{e1ecff}{6.0 $\pm$ 0.8} & \cellcolor[HTML]{e1ecff}{7.1 $\pm$ 2.4} \\
        \cellcolor[HTML]{e1ecff}{} & \cellcolor[HTML]{e1ecff}{\ding{55}} & \cellcolor[HTML]{e1ecff}{\ding{55}} & \cellcolor[HTML]{e1ecff}{\ding{51}} & \cellcolor[HTML]{e1ecff}{18.0 $\pm$ 4.2} & \cellcolor[HTML]{e1ecff}{20.9 $\pm$ 5.2} & \cellcolor[HTML]{e1ecff}{9.1 $\pm$ 2.1} & \cellcolor[HTML]{e1ecff}{14.4 $\pm$ 3.0} & \cellcolor[HTML]{e1ecff}{4.9 $\pm$ 0.6} & \cellcolor[HTML]{e1ecff}{5.9 $\pm$ 0.9} & \cellcolor[HTML]{e1ecff}{20.8 $\pm$ 4.9} & \cellcolor[HTML]{e1ecff}{24.1 $\pm$ 6.0} & \cellcolor[HTML]{e1ecff}{10.8 $\pm$ 2.5} & \cellcolor[HTML]{e1ecff}{17.0 $\pm$ 3.5} & \cellcolor[HTML]{e1ecff}{5.8 $\pm$ 0.7} & \cellcolor[HTML]{e1ecff}{7.0 $\pm$ 1.1} & \cellcolor[HTML]{e1ecff}{14.7 $\pm$ 3.4} & \cellcolor[HTML]{e1ecff}{17.0 $\pm$ 4.2} & \cellcolor[HTML]{e1ecff}{7.3 $\pm$ 1.7} & \cellcolor[HTML]{e1ecff}{11.5 $\pm$ 2.4} & \cellcolor[HTML]{e1ecff}{3.9 $\pm$ 0.5} & \cellcolor[HTML]{e1ecff}{4.7 $\pm$ 0.7} \\
        \cellcolor[HTML]{e1ecff}{} & \cellcolor[HTML]{e1ecff}{\ding{55}} & \cellcolor[HTML]{e1ecff}{\ding{55}} & \cellcolor[HTML]{e1ecff}{\ding{55}} & \cellcolor[HTML]{e1ecff}{14.9 $\pm$ 1.1} & \cellcolor[HTML]{e1ecff}{19.5 $\pm$ 1.4} & \cellcolor[HTML]{e1ecff}{5.6 $\pm$ 0.9} & \cellcolor[HTML]{e1ecff}{7.3 $\pm$ 0.4} & \cellcolor[HTML]{e1ecff}{5.0 $\pm$ 0.8} & \cellcolor[HTML]{e1ecff}{6.5 $\pm$ 0.6} & \cellcolor[HTML]{e1ecff}{17.2 $\pm$ 1.3} & \cellcolor[HTML]{e1ecff}{22.5 $\pm$ 1.6} & \cellcolor[HTML]{e1ecff}{6.6 $\pm$ 1.1} & \cellcolor[HTML]{e1ecff}{10.2 $\pm$ 0.5} & \cellcolor[HTML]{e1ecff}{5.9 $\pm$ 0.9} & \cellcolor[HTML]{e1ecff}{7.7 $\pm$ 0.7} & \cellcolor[HTML]{e1ecff}{12.1 $\pm$ 0.9} & \cellcolor[HTML]{e1ecff}{15.8 $\pm$ 1.1} & \cellcolor[HTML]{e1ecff}{4.5 $\pm$ 0.7} & \cellcolor[HTML]{e1ecff}{5.0 $\pm$ 0.3} & \cellcolor[HTML]{e1ecff}{4.0 $\pm$ 0.6} & \cellcolor[HTML]{e1ecff}{5.2 $\pm$ 0.5} \\
        \bottomrule
        \end{tabular}
    }
    \label{tab:rq4_new}
\end{table*}

\noindent
\textbf{Experimental Design.}
We conduct ablation studies across two dimensions: \textit{component analysis} and \textit{model robustness evaluation}.
We evaluate configurations removing the APR (\mytitle-no-APR), 
the GUI feedback (\mytitle-no-gui), 
both agents (\mytitle-no-apr-no-gui), the ContextSearchTool (\mytitle-no-context), and all agentic components (\mytitle-no-agent).

\noindent
\textbf{Results.}
Tables~\ref{tab:rq4_models} and~\ref{tab:rq4_new} present cross-language performance across Python, JavaScript, and TypeScript.
JavaScript consistently outperforms Python, with leading models achieving 44.0\% Exec@3 (Qwen3-Coder) and 44.5\% Exec@3 (Claude-Sonnet-4) compared to 38.1\% and 36.8\% respectively in Python.
TypeScript exhibits systematic degradation, with Qwen3-Coder reaching 31.0\% Exec@3 and Claude-Sonnet-4 achieving 30.1\% Exec@3, approximately 20\% lower than Python baselines.
These cross-language patterns remain stable across model tiers: intermediate models (GPT-5, GPT-4o, GPT-5-mini) demonstrate 14-16\% JavaScript improvements and 15-25\% TypeScript reductions, while smaller models (GPT-4o-mini, Grok-3-mini) maintain similar relative performance gaps.
Ablation studies in Table~\ref{tab:rq4_new} reveal consistent component criticality across languages.
For GPT-5-mini, APR removal causes 8.5pp Exec@3 degradation in Python (26.8\% to 18.3\%), 9.8pp in JavaScript (30.9\% to 21.1\%), and 6.9pp in TypeScript (21.8\% to 14.9\%).
GUI feedback elimination reduces Play@3 by 4.5pp in Python, 4.7pp in JavaScript, and 4.2pp in TypeScript, confirming its universal importance for behavioral validation.
Higher-capability models exhibit amplified degradation: Claude-Sonnet-3.7 loses 14.1pp Python Exec@3 without APR, with proportional losses across JavaScript (16.4pp) and TypeScript (11.5pp).
Complete component removal (no APR, GUI feedback, context) causes catastrophic performance collapse, validating the synergistic architecture of \mytitle.

\begin{table*}[htbp!]
    \centering
    \vspace{-0.2cm}
    \caption{Performance (\%) of \mytitle across different LLMs, 5 runs + 95\% CI (t-distribution).}
    \resizebox{\linewidth}{!}
    {
        \begin{tabular}{l|cccccc|cccccc|cccccc}
        \toprule
        \multirow{2}{*}{\textbf{LLMs}} & \multicolumn{6}{c|}{\textbf{Python}} & \multicolumn{6}{c|}{\textbf{JavaScript}} & \multicolumn{6}{c}{\textbf{TypeScript}} \\
        \cmidrule(lr){2-7} \cmidrule(lr){8-13} \cmidrule(lr){14-19}
        & \textbf{Exec@1} & \textbf{Exec@3} & \textbf{Pass@1} & \textbf{Pass@3} & \textbf{Play@1} & \textbf{Play@3} & \textbf{Exec@1} & \textbf{Exec@3} & \textbf{Pass@1} & \textbf{Pass@3} & \textbf{Play@1} & \textbf{Play@3} & \textbf{Exec@1} & \textbf{Exec@3} & \textbf{Pass@1} & \textbf{Pass@3} & \textbf{Play@1} & \textbf{Play@3} \\
        \midrule
        \cellcolor[HTML]{d9ead3}{Qwen3-Coder} & \cellcolor[HTML]{d9ead3}{\textbf{32.4 $\pm$ 5.8}} & \cellcolor[HTML]{d9ead3}{\textbf{38.1 $\pm$ 4.9}} & \cellcolor[HTML]{d9ead3}{17.6 $\pm$ 2.1} & \cellcolor[HTML]{d9ead3}{22.0 $\pm$ 3.2} & \cellcolor[HTML]{d9ead3}{16.0 $\pm$ 1.1} & \cellcolor[HTML]{d9ead3}{18.9 $\pm$ 4.0} & \cellcolor[HTML]{d9ead3}{\textbf{37.4 $\pm$ 6.7}} & \cellcolor[HTML]{d9ead3}{44.0 $\pm$ 5.7} & \cellcolor[HTML]{d9ead3}{20.8 $\pm$ 2.5} & \cellcolor[HTML]{d9ead3}{26.0 $\pm$ 3.8} & \cellcolor[HTML]{d9ead3}{18.9 $\pm$ 1.3} & \cellcolor[HTML]{d9ead3}{22.4 $\pm$ 4.7} & \cellcolor[HTML]{d9ead3}\textbf{26.4 $\pm$ 4.7} & \cellcolor[HTML]{d9ead3}\textbf{31.0 $\pm$ 4.0} & \cellcolor[HTML]{d9ead3}{14.1 $\pm$ 1.7} & \cellcolor[HTML]{d9ead3}{17.6 $\pm$ 2.6} & \cellcolor[HTML]{d9ead3}{12.8 $\pm$ 0.9} & \cellcolor[HTML]{d9ead3}{15.1 $\pm$ 3.2} \\
        GPT-5 & 25.4 $\pm$ 5.9 & 29.5 $\pm$ 4.0 & 14.5 $\pm$ 2.5 & 19.1 $\pm$ 1.0 & 10.9 $\pm$ 0.7 & 12.4 $\pm$ 1.8 & 29.0 $\pm$ 5.9 & 33.7 $\pm$ 4.0 & 16.5 $\pm$ 2.5 & 21.8 $\pm$ 1.0 & 12.4 $\pm$ 0.7 & 14.2 $\pm$ 1.8 & 19.6 $\pm$ 4.6 & 22.8 $\pm$ 3.1 & 11.2 $\pm$ 1.9 & 14.8 $\pm$ 0.8 & 8.4 $\pm$ 0.5 & 9.6 $\pm$ 1.4 \\
        \cellcolor[HTML]{d9ead3}{Claude-Sonnet-4} & \cellcolor[HTML]{d9ead3}{30.4 $\pm$ 1.5} & \cellcolor[HTML]{d9ead3}{36.8 $\pm$ 7.1} & \cellcolor[HTML]{d9ead3}{\textbf{18.0 $\pm$ 1.5}} & \cellcolor[HTML]{d9ead3}{\textbf{22.9 $\pm$ 2.1}} & \cellcolor[HTML]{d9ead3}{\textbf{17.1 $\pm$ 1.8}} & \cellcolor[HTML]{d9ead3}{\textbf{20.3 $\pm$ 2.5}} & \cellcolor[HTML]{d9ead3}{36.8 $\pm$ 1.8} & \cellcolor[HTML]{d9ead3}\textbf{44.5 $\pm$ 3.6} & \cellcolor[HTML]{d9ead3}\textbf{21.8 $\pm$ 1.8} & \cellcolor[HTML]{d9ead3}\textbf{27.7 $\pm$ 2.5} & \cellcolor[HTML]{d9ead3}\textbf{20.7 $\pm$ 2.2} & \cellcolor[HTML]{d9ead3}\textbf{24.6 $\pm$ 3.0} & \cellcolor[HTML]{d9ead3}{24.9 $\pm$ 1.2} & \cellcolor[HTML]{d9ead3}{30.1 $\pm$ 5.8} & \cellcolor[HTML]{d9ead3}\textbf{14.7 $\pm$ 1.2} & \cellcolor[HTML]{d9ead3}\textbf{18.7 $\pm$ 1.7} & \cellcolor[HTML]{d9ead3}\textbf{14.0 $\pm$ 1.5} & \cellcolor[HTML]{d9ead3}\textbf{16.6 $\pm$ 2.0} \\
        Claude-Sonnet-3.7 & {29.5 $\pm$ 2.7} & {35.6 $\pm$ 4.4} & {15.5 $\pm$ 0.6} & {19.4 $\pm$ 2.9} & {13.9 $\pm$ 0.9} & {17.4 $\pm$ 1.3} & {36.4 $\pm$ 3.3} & {43.9 $\pm$ 5.3} & {19.1 $\pm$ 0.7} & {23.9 $\pm$ 3.5} & {17.2 $\pm$ 1.1} & {21.5 $\pm$ 1.6} & {22.4 $\pm$ 2.0} & {27.0 $\pm$ 3.3} & {11.8 $\pm$ 0.5} & {14.7 $\pm$ 2.2} & {10.5 $\pm$ 0.7} & {13.2 $\pm$ 1.0} \\
        \cellcolor[HTML]{d9ead3}{GPT-4o} & \cellcolor[HTML]{d9ead3}{23.6 $\pm$ 3.2} & \cellcolor[HTML]{d9ead3}{27.1 $\pm$ 2.8} & \cellcolor[HTML]{d9ead3}{14.1 $\pm$ 0.3} & \cellcolor[HTML]{d9ead3}{18.2 $\pm$ 1.6} & \cellcolor[HTML]{d9ead3}{9.4 $\pm$ 0.8} & \cellcolor[HTML]{d9ead3}{13.6 $\pm$ 2.8} & \cellcolor[HTML]{d9ead3}{27.5 $\pm$ 3.7} & \cellcolor[HTML]{d9ead3}{31.5 $\pm$ 3.3} & \cellcolor[HTML]{d9ead3}{16.4 $\pm$ 0.4} & \cellcolor[HTML]{d9ead3}{21.2 $\pm$ 1.9} & \cellcolor[HTML]{d9ead3}{10.9 $\pm$ 0.9} & \cellcolor[HTML]{d9ead3}{15.8 $\pm$ 3.3} & \cellcolor[HTML]{d9ead3}{17.8 $\pm$ 2.4} & \cellcolor[HTML]{d9ead3}{20.4 $\pm$ 2.1} & \cellcolor[HTML]{d9ead3}{10.6 $\pm$ 0.2} & \cellcolor[HTML]{d9ead3}{13.7 $\pm$ 1.2} & \cellcolor[HTML]{d9ead3}{7.1 $\pm$ 0.6} & \cellcolor[HTML]{d9ead3}{10.2 $\pm$ 2.1} \\
        GPT-5-mini & 23.0 $\pm$ 1.4 & 26.8 $\pm$ 2.4 & 12.7 $\pm$ 1.1 & 19.7 $\pm$ 2.1 & 8.3 $\pm$ 1.3 & 9.8 $\pm$ 2.1 & {26.1 $\pm$ 1.6} & {30.4 $\pm$ 2.8} & {14.4 $\pm$ 1.3} & {22.3 $\pm$ 2.5} & {9.4 $\pm$ 0.6} & {11.1 $\pm$ 2.5} & {18.4 $\pm$ 1.1} & {21.5 $\pm$ 1.9} & {10.2 $\pm$ 0.9} & {15.8 $\pm$ 1.7} & {6.6 $\pm$ 0.4} & {7.8 $\pm$ 1.7} \\
        \cellcolor[HTML]{d9ead3}{GLM-4.5} & \cellcolor[HTML]{d9ead3}{21.8 $\pm$ 2.5} & \cellcolor[HTML]{d9ead3}{24.8 $\pm$ 3.4} & \cellcolor[HTML]{d9ead3}{8.0 $\pm$ 1.4} & \cellcolor[HTML]{d9ead3}{11.9 $\pm$ 1.7} & \cellcolor[HTML]{d9ead3}{7.6 $\pm$ 1.0} & \cellcolor[HTML]{d9ead3}{9.6 $\pm$ 0.9} & \cellcolor[HTML]{d9ead3}{24.1 $\pm$ 2.9} & \cellcolor[HTML]{d9ead3}{27.4 $\pm$ 4.0} & \cellcolor[HTML]{d9ead3}{8.8 $\pm$ 1.6} & \cellcolor[HTML]{d9ead3}{13.1 $\pm$ 2.0} & \cellcolor[HTML]{d9ead3}{8.4 $\pm$ 1.2} & \cellcolor[HTML]{d9ead3}{10.6 $\pm$ 1.1} & \cellcolor[HTML]{d9ead3}{16.8 $\pm$ 1.9} & \cellcolor[HTML]{d9ead3}{19.1 $\pm$ 2.6} & \cellcolor[HTML]{d9ead3}{6.2 $\pm$ 1.1} & \cellcolor[HTML]{d9ead3}{9.2 $\pm$ 1.3} & \cellcolor[HTML]{d9ead3}{5.9 $\pm$ 0.8} & \cellcolor[HTML]{d9ead3}{7.4 $\pm$ 0.7} \\
        DeepSeek-V3 & 19.2 $\pm$ 1.4 & 23.3 $\pm$ 2.0 & 9.8 $\pm$ 0.5 & 14.5 $\pm$ 3.7 & 7.5 $\pm$ 0.8 & 9.9 $\pm$ 1.1 & 23.0 $\pm$ 1.6 & 27.9 $\pm$ 2.4 & 11.7 $\pm$ 0.6 & 17.4 $\pm$ 4.4 & 9.0 $\pm$ 1.0 & 11.9 $\pm$ 1.3 & 15.4 $\pm$ 1.1 & 18.7 $\pm$ 1.6 & 7.9 $\pm$ 0.4 & 11.7 $\pm$ 3.0 & 6.0 $\pm$ 0.6 & 8.0 $\pm$ 0.9 \\
        \cellcolor[HTML]{d9ead3}{GPT-4o-mini} & \cellcolor[HTML]{d9ead3}{15.5 $\pm$ 2.0} & \cellcolor[HTML]{d9ead3}{19.0 $\pm$ 1.1} & \cellcolor[HTML]{d9ead3}{7.8 $\pm$ 0.8} & \cellcolor[HTML]{d9ead3}{11.0 $\pm$ 1.6} & \cellcolor[HTML]{d9ead3}{5.1 $\pm$ 0.5} & \cellcolor[HTML]{d9ead3}{7.8 $\pm$ 0.6} & \cellcolor[HTML]{d9ead3}{17.6 $\pm$ 2.3} & \cellcolor[HTML]{d9ead3}{21.5 $\pm$ 1.3} & \cellcolor[HTML]{d9ead3}{8.8 $\pm$ 0.9} & \cellcolor[HTML]{d9ead3}{12.5 $\pm$ 1.8} & \cellcolor[HTML]{d9ead3}{5.8 $\pm$ 0.6} & \cellcolor[HTML]{d9ead3}{8.8 $\pm$ 0.7} & \cellcolor[HTML]{d9ead3}{12.5 $\pm$ 1.6} & \cellcolor[HTML]{d9ead3}{15.4 $\pm$ 0.9} & \cellcolor[HTML]{d9ead3}{6.3 $\pm$ 0.6} & \cellcolor[HTML]{d9ead3}{8.9 $\pm$ 1.3} & \cellcolor[HTML]{d9ead3}{4.1 $\pm$ 0.4} & \cellcolor[HTML]{d9ead3}{6.3 $\pm$ 0.5} \\
        Grok-3-mini & 13.5 $\pm$ 1.4 & 16.0 $\pm$ 2.8 & 4.8 $\pm$ 0.6 & 7.8 $\pm$ 1.2 & 4.7 $\pm$ 0.6 & 6.3 $\pm$ 0.6 & 16.5 $\pm$ 1.7 & 19.5 $\pm$ 3.4 & 5.9 $\pm$ 0.7 & 9.5 $\pm$ 1.5 & 5.7 $\pm$ 0.7 & 7.7 $\pm$ 0.7 & 10.1 $\pm$ 1.1 & 12.0 $\pm$ 2.1 & 3.6 $\pm$ 0.5 & 5.9 $\pm$ 0.9 & 1.5 $\pm$ 0.5 & 4.7 $\pm$ 0.5 \\
        \bottomrule
        \end{tabular}
    }
    \vspace{-0.3cm}
    \label{tab:rq4_models}
\end{table*}

\intuition{
{\bf Answer to RQ-3}: The ablation study reveals that both APR and GUI feedback make essential contributions to \mytitle's effectiveness. 
The model robustness evaluation demonstrates that \mytitle provides consistent improvements across diverse LLM architectures.
}

\label{sec:rq3}

\subsection{Case Study}

To demonstrate practical capabilities, we present a case study examining its multi-modal reasoning and adaptive testing strategies on a representative 2048 game implementation. This case study qualitatively showcases \mytitle's advanced capacity for deep, context-aware reasoning and its ability to transform complex visual game states into structured testing protocols while maintaining strategic gameplay coherence. As shown in Figure~\ref{fig:2048_tencent}, the scenario involves \mytitle{} analyzing an early-game 2048 state. The case study demonstrates three aspects of \mytitle{}'s operation:

\begin{figure}[htbp]
    \centering
    \includegraphics[width=\textwidth]{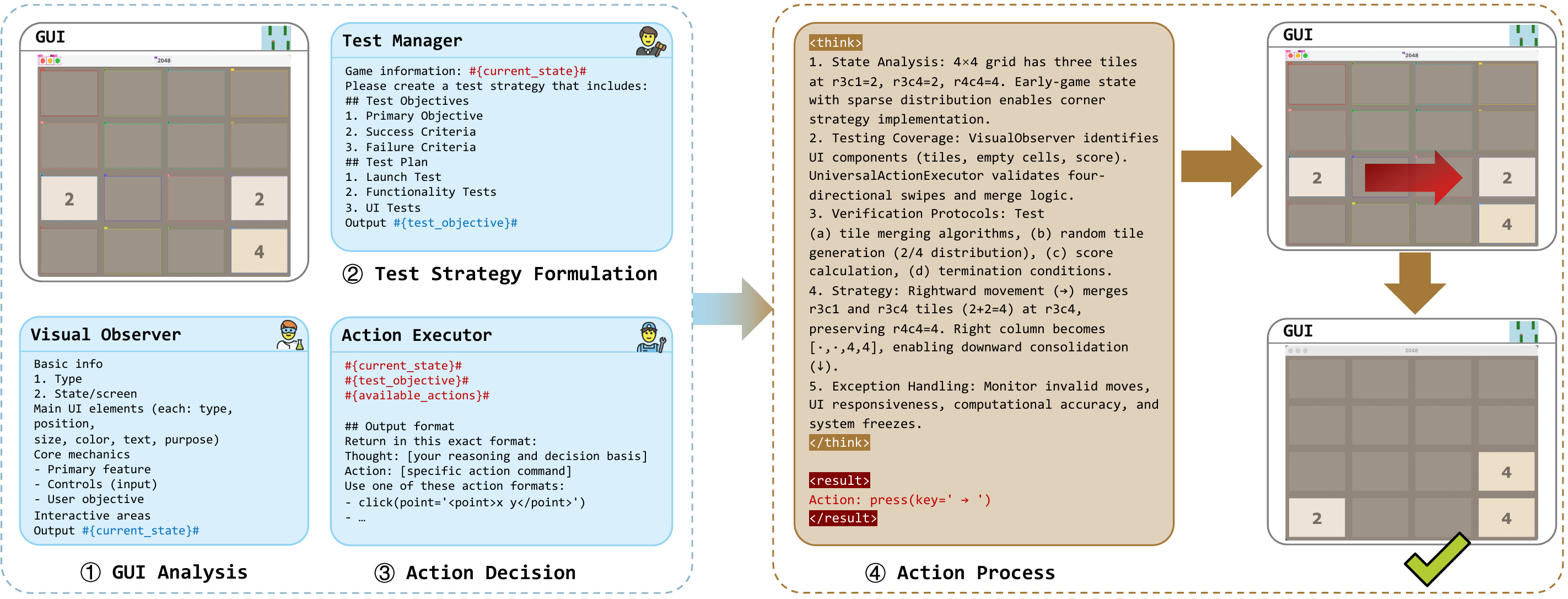}
    \caption{Case study of \mytitle{} testing a 2048 game.}
    \label{fig:2048_tencent}
\end{figure}

\noindent\textbf{Multi-Phase Reasoning Architecture.} \mytitle{} employs a three-phase reasoning process that integrates testing objectives with gameplay optimization. First, the \texttt{VisualObserver} component captures and analyzes the current game state, identifying tile positions (r3c1=2, r3c4=2, r4c4=4) and recognizing the sparse early-game configuration. Second, the system constructs a testing protocol encompassing functional validation (merge algorithms, score computation, random tile generation), performance assessment (UI responsiveness), and boundary condition testing (invalid moves, termination detection). Third, \mytitle{} synthesizes these analyses to select actions that simultaneously advance strategic gameplay while maximizing test coverage.

\noindent\textbf{Adaptive Test Strategy Generation.} 
Unlike static testing approaches that rely on predefined test sequences, \mytitle{} dynamically generates testing strategies based on real-time visual analysis. 
In the depicted scenario, the system recognizes that a rightward swipe (→) operation serves dual objectives: 
(1) implementing corner strategy optimization by merging the two 2-tiles into a 4-tile at position r3c4, 
and (2) validating critical game mechanics including swipe responsiveness, tile merger algorithms, and score calculation accuracy. 
This approach ensures comprehensive functionality validation while maintaining viable gameplay progression, 
addressing a key limitation of conventional GUI testing frameworks.

\noindent\textbf{Exception-Aware Validation Protocol.} The case study highlights \mytitle's proactive approach to anomaly detection and edge case handling. The system explicitly considers failure scenarios including invalid move detection, UI freezing, computational accuracy errors, and improper termination conditions. This comprehensive exception handling demonstrates the framework's ability to identify behavioral failures that would be missed by compilation-based testing or unit test approaches, directly addressing the semantic correctness gaps identified in existing benchmarks. The presented case study exemplifies how \mytitle{} transcends traditional testing limitations by integrating strategic reasoning with systematic validation. The system's ability to process visual information, generate structured testing protocols, and execute dual-objective actions (gameplay optimization + functionality validation) represents a significant advancement in automated GUI testing for interactive applications. This approach provides a robust foundation for detecting and repairing the silent logic flaws that characterize GUI application failures, thereby establishing behavioral correctness as a fundamental requirement for reliable code generation.

\revise{\noindent\textbf{Success Case (2048).} MetaGPT generated a version where the tile font color was white on a white background, making numbers invisible. Standard unit tests passed, but the game was unplayable. PlayTester detected a lack of visible changes and fixed the rendering code.}

\revise{\noindent\textbf{Failed Case.} In scenarios requiring high-frequency feedback (e.g., a bug appearing only when running at $>60$ FPS) or long-duration survival (e.g., a crash occurring only after 2 minutes of gameplay in Flappy Bird), all baseline methods, including PlayCoder, failed. This failure mode is observed in 4 out of the 43 projects in \datasetname{} (roughly 9\%), concentrated in fast-paced arcade games (Flappy Bird, Snake) and real-time physics simulations. We regard this as a limitation of behavioral testing under discrete polling, compounded by GPU compute bottlenecks that keep end-to-end inference latency from meeting the tight timing budgets of these scenarios; we leave higher-frequency visual sampling and lower-latency inference to future work.}

\label{sec:case_studies}

\vspace{-0.3cm}
\section{Related Work}
\subsection{Code Generation with LLMs}

Advances in pre-training have substantially improved neural code generation for both academia and industry~\cite{li2022competition,shen2022incorporating,nijkamp2023codegen,fried2023incoder}.
This progress has produced a rich family of large language models with strong coding ability~\cite{openai2022chatgpt,wei2024magicoder,roziere2023code,bai2023qwen,deepseekcoder,zheng2024opencodeinterpreter,yin2024rectifier,yin2024you,xia2024automated}, and subsequent work continues to scale model quality while also synthesizing the landscape through surveys~\cite{liu2024deepseek,deepseekr1,wang2021codet5,li2026advances,wang2025towards}.
Overall, the field has moved from comparatively shallow token prediction toward models that better capture complex structure and cross-file dependencies.

To adapt LLMs to diverse generation settings, a large body of work studies prompt engineering and introduces reusable interaction patterns, including Structured Chain-of-Thought~\cite{yin2024thinkrepair,li2025structured}, Self-planning~\cite{jiang2024self}, Self-debug~\cite{chen2023teaching,xia2024automated}, and Self-collaboration~\cite{dong2024self,yin2024rectifier}.
Beyond generic prompting, repository-level methods make generation sensitive to project context: A3-CodGen~\cite{A3CodGen} explicitly models local, global, and third-party library usage to support code reuse, while Shrivastava et al.~\cite{shrivastava2023repository} develop repository-aware prompt construction strategies.
Most relevant to long-horizon completion, Hierarchical Context Pruning (HCP)~\cite{zhang2025hierarchical} assembles prompts that respect topological dependency order and prune non-essential global and implementation detail, improving CrossCodeEval accuracy on five of six repository-trained code LLMs while increasing throughput.
SolEval~\cite{peng2025soleval} introduces the first repository-level benchmark for Solidity code generation.
PrefGen~\cite{peng2025preference} builds on SolEval within a preference-driven training and evaluation framework, fine-tuning LLMs with supervised fine-tuning (SFT) and direct preference optimization (DPO) for repository-level Solidity generation.
Parallel to these modeling ideas, agentic code generation frameworks decompose software work into coordinated procedures.
MetaGPT~\cite{hong2024metagpt} simulates a software organization via specialized roles; OpenManus~\cite{openmanus2025} tackles complex tasks with a multi-agent execution stack; DeepCode~\cite{deepcode2025} focuses on reliable automation for common development routines.

Beyond agents, recent benchmarks increasingly emphasize pragmatic coding competence~\cite{chen2021evaluating,hendrycks2021measuring,peng2026repogenesis}.
LiveCodeBench~\cite{jain2025livecodebench} reduces evaluation leakage by continuously refreshing problem sets.
CoderEval~\cite{yu2024codereval} foregrounds realistic coding scenarios, and Evocodebench~\cite{li2024evocodebench} stresses generation inside practical software projects.
ClassEval~\cite{du2024evaluating} shifts attention from isolated functions to class-level synthesis.
Finally, SWE-Bench~\cite{jimenez2024swe} measures GitHub issue resolution and has catalyzed a wave of follow-on studies on practical software engineering tasks~\cite{tao2024magis,zan2025multi,yang2025swebench,guo2025omnigirl,zhang2025swe,he2025swe,aleithan2025revisiting,aleithan2024swe,xie2025swe,li2025fea}.
However, these benchmarks largely omit behavioral correctness for generated \emph{GUI} applications.
Moreover, mainstream generation stacks still emphasize compile- and test-oriented functional signals rather than grounding iteration in visual, interactive execution feedback.

\subsection{GUI Interaction Understanding, Testing, and Generation}

Traditional GUI automation relies on \emph{rule-based} exploration.
Random fuzzers such as Android Monkey exercise apps with pseudo-random inputs but lack systematic coverage and semantic checks~\cite{android_monkey}.
Model- and search-based testing improves exploration: Dynodroid adds system awareness to input generation~\cite{dynodroid_fse13}; Sapienz jointly optimizes coverage, fault detection, and test suite size with multi-objective search~\cite{sapienz_issta16}; Stoat learns stochastic state models for event sequences~\cite{stoat_icsme17}.
Earlier frameworks also compile event-flow graphs and widget hierarchies into scalable tests and oracles~\cite{memon_tse03}.
Collectively, these methods emphasize structural models and coverage more than validating \emph{semantic} GUI behavior against natural-language intent. 
On the other hand, learning-based methods broaden GUI \emph{understanding} and downstream automation.
UIED~\cite{xie2020uied} hybridizes CV and ML for cross-platform element detection; Screen Recognition~\cite{zhang2021screen} infers accessibility metadata from pixels; Owl Eyes~\cite{liu2020owl} flags display defects visually.
Deep GUI~\cite{yazdanibanafshedaragh2021deep}, ResPlay~\cite{zhang2023resplay}, and Baral et al.~\cite{baral2024automating} turn strong perception into black-box inputs, cross-platform record-and-replay, and mobile test oracles, respectively.
Mansur et al.~\cite{mansur2023aidui} further target UX risks via dark-pattern detection.
Rico~\cite{deka_rico_uist17} supplies large-scale layouts for data-driven modeling; ScreenAI~\cite{screenai_arxiv24} improves widget recognition, captioning, and instruction following on screens.
WebArena and Mind2Web~\cite{webarena_neurips23,mind2web_neurips23} benchmark multi-step web interaction.
These resources support screen understanding and action planning, but not repository-aware synthesis and repair of full GUI application code. On \emph{GUI generation and testing}, \textit{pix2code} and \textit{web2code}~\cite{beltramelli_pix2code_2017,yun2024web2code} translate designs into code; \textit{Seq2Act}~\cite{li_seq2act_kdd20} maps language to UI action traces; GPTDroid~\cite{liu2024make} and Humanoid~\cite{li2019humanoid} drive mobile exploration with LLMs or deep policies.
Such systems mainly yield layouts, scripts, or action loops with limited guarantees on end-to-end runtime logic, a gap also reflected in benchmarks for automated GUI testing~\cite{chen2024gui,baral2024automating}.
Unlike coverage- or crash-oriented testers~\cite{dynodroid_fse13,sapienz_issta16,stoat_icsme17} and agents scored on completing tasks over existing GUIs~\cite{webarena_neurips23,mind2web_neurips23,screenai_arxiv24},
our approach feeds dynamic execution feedback into the code generation loop to detect and repair logic errors (e.g., rule violations in GUI games) that compile- and run-based checks fail to capture.

\section{Threats to Validity}

\noindent\textbf{External Threats.} \mytitle's effectiveness is inherently constrained by the capabilities of underlying vision-language models. 
Current VLMs exhibit limitations in recognizing fine-grained GUI elements and interpreting complex visual semantics, 
which directly translate to constraints in our testing and validation capabilities. 
This threat will be mitigated by the evolution vision-language models in the future. 
Furthermore, the probabilistic nature of LLMs introduces inherent instability in complex contexts, 
causing response variations even if the temperature is set to zero. 
We mitigate this by conducting all experiments with 5 repetitions to ensure stable and reliable results.

\noindent\textbf{Internal Threats.} GUI-based behavioral testing relies on screenshot-based analysis, 
unable to capture every critical frame in dynamic applications (e.g., FPS games), 
potentially missing bug detection opportunities in unfavorable timing scenarios. 
Furthermore, the context retrieval mechanisms may face scalability challenges in large, 
complex repositories with extensive dependencies.
The benchmark scenarios and number of applications evaluated may be limited in scope.
We address this limitation by selecting applications that span multiple domains and platforms.

\section{Conclusion}

This paper addresses limitations in evaluating and generating GUI application code. 
We identify that existing benchmarks fail to capture behavioral correctness in interactive applications, 
where syntactically correct code can exhibit catastrophic logic flaws that traditional unit tests miss.
We introduce \mytitle, a novel multi-agent framework that integrates automated GUI testing with iterative program repair. 
Our approach employs PlayTester to simulate user interactions and detect behavioral deviations, 
paired with PlayRefiner that autonomously debugs and refines the generated code. 
Through systematic evaluation on \revise{diverse multilingual (Python, TypeScript, and JavaScript) GUI applications,} 
we demonstrated that this collaborative framework significantly outperforms baseline methods, 
achieving higher rates of functional correctness and semantic alignment.
Our work highlights the importance of runtime verification in complex code generation tasks 
and establishes a new paradigm for ensuring behavioral fidelity in interactive applications. 

\section{Data Availability}
The replication package of this paper \revise{(including prompts, code, and datasets)} is publicly available at \url{https://github.com/Tencent/PlayCoder}.

\section*{Acknowledgements}
This work was supported in part by National Key
Research and Development Program of China under Grant 2024YFB2705300, in part by the Shanghai Science and Technology Innovation Action Plan under Grant 23511100400.

\balance
\bibliographystyle{ACM-Reference-Format}
\bibliography{main}

\end{document}